\documentclass[10pt, journal, letterpaper, compsoc]{IEEEtran}

\usepackage[sort, compress, nobreak]{cite}

\usepackage[utf8]{inputenc}
\usepackage[english]{babel}
\usepackage{amsfonts, amssymb, amsmath, amsthm, gensymb, mathcomp, stmaryrd}
\usepackage{url}
\usepackage{graphicx}
\usepackage{multirow}
\usepackage{rotating}
\usepackage{array}
\usepackage[ruled, noend, linesnumbered]{algorithm2e}
\usepackage[obeyDraft]{todonotes}
\usepackage{subfig}

\usepackage{xcolor}

\usepackage{tikz}
\usetikzlibrary{calc}
\usetikzlibrary{arrows}
\usetikzlibrary{positioning}
\usetikzlibrary{matrix}
\usepackage{pgfplots}
\pgfplotsset{width=7cm,compat=1.4}

\newcommand\cmod[1]{~\operatorname{cmod}~#1}
\newcommand\dist[2]{\operatorname{dist}_#1(#2)}

\newcommand\alert[1]{#1}

\newtheorem{definition}{Definition}
\newtheorem{theorem}{Theorem}
\newtheorem{property}{Property}

\begin{document}

\title{\alert{GPU-accelerated generation of \\correctly-rounded
    elementary functions}}

\author{Pierre Fortin, Mourad Gouicem and Stef Graillat
  \thanks{Authors are with UPMC Univ Paris 06 and CNRS UMR 7606, LIP6}
  \thanks{Address : 4 place Jussieu, F-75252, Paris cedex 05, France}
  \thanks{Contact: mourad.gouicem@lip6.fr} }

\maketitle

\begin{abstract}
  The IEEE 754-2008 standard recommends the correct rounding of \alert{some}
  elementary functions. This requires to solve the Table Maker's Dilemma which
  implies a huge amount of CPU computation time.  We consider in this paper
  accelerating such computations, namely Lefèvre algorithm on Graphics
  Processing Units (GPUs) which are massively parallel architectures with a
  partial SIMD execution (Single Instruction Multiple Data).

  We first propose an analysis of the Lefèvre {\it hard-to-round} argument
  search using the concept of continued fractions.  We then propose a new
  parallel search algorithm much more efficient on GPU thanks to its more
  regular control flow. We also present an efficient hybrid CPU-GPU deployment
  of the generation of \alert{the} polynomial approximations required in Lefèvre
  algorithm. In the end, we manage to obtain overall speedups up to 53.4x on one
  GPU over a sequential CPU execution, and up to 7.1x over a multi-core CPU,
  \alert{which enable a much faster solving of the Table Maker's Dilemma for the
    double precision format}.
\end{abstract}


\IEEEpeerreviewmaketitle

\section{Introduction}

\subsection{Problem}

The IEEE 754 standard specifies since 1985 the implementation of floating-point
operations in order to have portable and predictable numerical software. In its
latest revision in 2008 \cite{IEEE754_2008}, it defines formats (binary32,
binary64 and binary128), rounding modes (to the nearest and toward 0, $-\infty$
and $+\infty$) and operations $(+, -, \times, /, \sqrt{~})$ returning correctly
rounded values.

Furthermore, it recommends correct rounding of some elementary functions, like
\emph{log}, \emph{exp} and the trigonometric functions. As these functions are
transcendental, one cannot evaluate them exactly but have to approximate
them. However, it is hard to decide which intermediate precision is required to
guarantee a correctly rounded result -- the rounded evaluation of the
approximation must be equal to the rounded evaluation of the function with
infinite precision. This problem is known as the \emph{Table Maker's Dilemma} or
TMD \cite[chap. 12]{hfpa}.

\subsection{State of the art}

There exist theoretical bounds on the intermediate precision required for
correctly rounded functions \cite[chap. 12]{hfpa}, but these are not sharp
enough for efficient floating-point implementations of elementary functions. For
example the Nesterenko-Waldschmidt bound for the exponential in double precision
gives that $7.290.678$ bits of intermediate precision suffice to provide a
correctly rounded result. Hence {\it ad hoc} methods are needed to find a
sharper bound for each function.

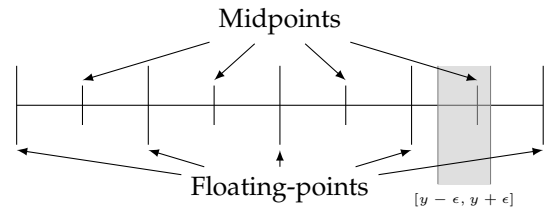
\begin{figure}
  \centering
  \begin{tikzpicture}[>=latex]
    \pgfmathsetmacro\intervalSize{7} 
    \pgfmathsetmacro\nbpts{4} 
    \pgfmathsetmacro\tangente{0.3} 

    \pgfmathsetmacro\bp{0} 
    \pgfmathsetmacro\ep{\intervalSize} 
    \pgfmathsetmacro\middle{(\bp+\ep)/2} 
    \pgfmathsetmacro\nbptsBis{\nbpts-1} 
    \pgfmathsetmacro\step{(\ep-\bp)/\nbpts}  
    \pgfmathsetmacro\ptsH{\tangente*\step} 
    \pgfmathsetmacro\midptsH{\ptsH/2} 
    \pgfmathsetmacro\shift{(((\ep-\bp)/2)*\ptsH)/\step}  

    \pgfmathsetmacro\PxRatio{0.4} 
    \pgfmathsetmacro\eps{0.2} 
    \pgfmathsetmacro\Px{(ceil(\nbptsBis))*\step + \PxRatio*\step} 
    \pgfmathsetmacro\PxMeps{\Px - \eps*\step} 
    \pgfmathsetmacro\PxPeps{\Px + \eps*\step} 

    \draw (\bp,0) -- (\ep,0);
    
    \foreach \x in {0,...,\nbpts}{ 
      \draw (\bp + \x*\step, -\ptsH) 
      -- (\bp + \x*\step, \ptsH); 
    }
    
    \foreach \x in {0,...,\nbptsBis}{ 
      \draw (\bp + \x*\step + 0.5*\step,-\midptsH) 
      -- (\bp + \x*\step + 0.5*\step, \midptsH); 
    }

    \fill [gray!50, opacity=0.5] 
    (\PxMeps, \ptsH) 
    rectangle (\PxPeps, -\shift);
    \node [minimum size=0, anchor=north] at (\Px, -\shift) {\tiny$[y-\epsilon , y + \epsilon]$};
    \draw [gray,thin] (\PxMeps,-\shift) 
    -- (\PxMeps, \ptsH) 
    (\PxPeps, \ptsH) 
    -- (\PxPeps,-\shift);

    \begin{scope}[yshift=2pt] 
      \node (breakpts) at (\middle, \shift) {Midpoints};
      \foreach \x in {0,...,\nbptsBis}{
        \draw [->]   (breakpts) 
        -- (\bp + \x*\step + 0.5*\step, \midptsH);
      }
    \end{scope}
    
    \begin{scope}[yshift=-2pt] 
      \node (fp) at (\middle, - \shift) {Floating-points};
      \foreach \x in {0,...,\nbpts}{
        \draw [->]   (fp) 
        -- (\bp + \x*\step, -\ptsH);
      }
    \end{scope}

  \end{tikzpicture}
  \caption{Example of undetermined correct rounding for rounding to nearest,
    where the rounding breakpoints are the midpoints of floating-point numbers.}
  \label{fig_hr_case}
\end{figure}

A first method introduced by Ziv \cite{Ziv91} was to compute an approximation
$y$ of a function value $f(x)$ with a bounded error of $\epsilon$ (containing
mathematical and round-off errors). As rounding modes are monotonic, if
$y-\epsilon$ and $y+\epsilon$ round to the same floating-point number, $f(x)$
does too : otherwise the correct rounding cannot be determined (see
Fig. \ref{fig_hr_case}). Hence having a correctly rounded result of $f(x)$ can
be done by refining the approximation $(y, \epsilon)$ -- decreasing $\epsilon$
-- until $y-\epsilon$ and $y+\epsilon$ round to the same floating-point
number. For the most common elementary functions, such an $\epsilon$ exists
according to the Lindemann–Weierstrass theorem when the function is evaluated at
almost all floating-point numbers \cite{Lindemann}.


However, the computation of many approximations can be avoided by precomputing
an $\epsilon$ guaranteeing correct-rounding of the evaluation of $f$ at any
floating-point number argument. This has to be done by finding the {\it
  hardest-to-round} arguments of the function, that is to say the arguments
requiring the highest precision to be correctly rounded when the function is
evaluated at. This precision guaranteeing the correct rounding for all arguments
is named the \emph{hardness-to-round of the function}. The {\it
  hardest-to-round} cases can be found by invoking Ziv algorithm at every
floating-point number in the domain of definition of the function, but this is
prohibitive ($O(2^p)$ when considering precision-$p$ floating point numbers as
arguments).

The first improvement was proposed by Lefèvre, Muller and Tisserand in
\cite{Lefevre1998} (Lefèvre algorithm). The main idea of their algorithm is to
split the domain of definition into several domains $D_i$, to ``isolate'' {\it
  hard-to-round} cases (HR-cases), and then to use Ziv algorithm to find the
{\it hardest-to-round} cases among them. This isolation is efficiently performed
using local affine approximations of the targeted function over $O(2^{2p/3})$
domains $D_i$. Stehlé, Lefèvre and Zimmermann extended this method in 2003
\cite{SLZ_journal} (SLZ algorithm) for higher degree approximations, using the
Coppersmith method for finding small roots of univariate modular equation over
$O(2^{p/2})$ domains $D_i$.

\subsection{Motivations and contributions}

Even if they are asymptotically and practically faster than exhaustive search,
Lefèvre and SLZ algorithms remain very computationally intensive. For example,
Lefèvre algorithm requires around five years of CPU time for the exponential
function over all double precision arguments, and the SLZ algorithm takes around
eight years of CPU time for the function $2^x$ over extended double precision
arguments in the interval $[1/2, 1]$\footnote{{SLZ} Algorithm - Results,
  \url{http://www.loria.fr/equipes/spaces/slz.en.html}}. Moreover, even if the
hardest-to-round cases of some functions in double precision are known
\cite[chap. 12.5]{hfpa}, it is still not the case for about half of the
univariate functions recommended by the IEEE standard 754-2008. Furthermore, we
still have no efficient way to find those of any elementary function in double
precision, and quadruple precision is out of reach. We will hence be interested
in accelerating the search of hardest-to-round case in double precision
(binary64).

As both algorithms split the domain of definition of the targeted function into
domains $D_i$ and search for HR-cases in them independently, these computations
are embarrassingly and massively parallel.  The purpose of this work is
therefore to accelerate these computations on Graphics Processing Units (GPUs),
which theoretically perform one order of magnitude better than CPUs thanks to
their massively parallel architectures.

We will focus here on Lefèvre algorithm, which has been used to generate all
known hardness-to-round in double precision \cite[chap. 12.5]{hfpa}. It is
asymptotically less efficient than SLZ as it considers more domains $D_i$
($O(2^{2p/3})$ against $O(2^{p/2})$). However, it performs less operations per
domain $D_i$ ($O(\log p)$ against $O(\operatorname{poly}(p))$). Therefore,
Lefèvre algorithm is as efficient as SLZ in practice for finding the
hardness-to-round of elementary functions for double precision format
\cite{dinechin2011} \cite[chap. 12]{hfpa} and offers fine-grained parallelism,
making it suitable for GPU.

In \cite{Gouicem12}, we discussed implementation techniques to deploy the
original Lefèvre algorithm efficiently on GPU which led to an average speedup of
15.4x with respect to the reference CPU implementation on one CPU core. The
major bottleneck of this GPU deployment was the control flow divergence which is
penalizing considering the partial SIMD execution (Single Instruction Multiple
Data) of the GPU. Hardware \cite{Collange12} and software \cite{Frey12, Han11}
general solutions have been proposed recently to address this problem on
GPU. However, these solutions are not efficient in our context as we have a very
fine computation grain for each GPU thread. Hence we here focus on algorithmic
solutions to tackle directly the origin of this divergence issue.

In this paper, we thus redesign Lefèvre algorithm with the continued fraction
formalism, which enables us to get a better understanding of it and to propose a
much more regular algorithm for searching HR-cases. More precisely, we strongly
reduce two major sources of divergence of Lefèvre algorithm: loop divergence and
branch divergence.  We also propose an efficient hybrid CPU-GPU deployment of
the generation of polynomial approximations $D_i$ using fixed multi-precision
operations on GPU.  These contributions enable on GPU an overall speedup of
53.4x over Lefèvre's original sequential CPU implementation, and of 7.1x over
six CPU cores (with two-way SMT). Finally, as we obtain in the end the same
HR-cases as Lefèvre, de Dinechin and Muller experiments
\cite{dinechin2011,Lefevre_hrcases} we also strengthen the confidence in the
generated HR-cases.

\subsection{Outline}

We will first introduce some notions on GPU architecture and divergence in Sect.
\ref{sect_gpu}. Then we will present in Sect. \ref{sect_math_pre} some
mathematical background on the Table Maker's Dilemma and properties of the set
\mbox{$\{a\cdot x \mod{1}~|~x<n \}$} with $a$ fixed. In
Sect. \ref{sect_hr_search}, we will detail the HR-case search step of Lefèvre
algorithm and of the new and more regular algorithm. In the same
Sect. \ref{sect_hr_search}, we will also present their deployment on GPU. In
Sect. \ref{sect_pol_approx} we will detail how to efficiently generate on GPU
the polynomial approximations $D_i$ needed by the two HR-case searches. And
finally, we will present performance results in Sect. \ref{sect_perf} and
conclude in Sect.  \ref{sect_ccl}.


\section{GPU computing}
\label{sect_gpu}

Graphics Processing Units (GPUs) are many-core devices originally intended for
graphics computations. However since mid-2000s they became increasingly used for
high performance scientific computing since their massively parallel
architectures theoretically perform one order of magnitude better than CPUs, and
since general-purpose languages adapted to GPUs like CUDA
\cite{nvidia_best_practice} and OpenCL \cite{OpenCL} have emerged.  In this
section we briefly describe the architecture of the NVIDIA GPU used to test our
deployments (the Fermi architecture), GPU programming in CUDA and the divergence
problems arising from the partial SIMD execution on GPU. We use here the CUDA
nomenclature.

\subsection{GPU architecture and CUDA programming}

From a hardware point of view, a GPU is composed of several {\it Streaming
  Multiprocessors} denoted SM (14 on Fermi C2070), each being a SIMD unit
(Single Instruction Multiple Data) \cite{nvidia_programming_guide}.  A SM is
composed of multiple execution units or {\it CUDA cores} (32 on Fermi) sharing
the same pipeline and many registers (32768 on Fermi).  GPU memory is organized
in two levels: {\it device memory}, which can be accessed by any SM on the
device; and {\it shared memory}, which is local to each SM. The device memory
accesses are cached on the Fermi architecture.

From a software point of view, the developer writes in CUDA a scalar code for
one function designed to be executed on the device, namely a {\it kernel}. At
runtime, many {\it threads} are created by {\it blocks} and bundled into a {\it
  grid} to run the same kernel concurrently on the device. Each block is
assigned to a SM. Within each block, threads are executed by groups of 32 called
{\it warps}. The ratio of the number of resident warps (number of warps a SM can
process at the same time) to the maximum number of resident warps per SM is
named the {\it occupancy}. In order to increase the occupancy the number of
blocks and their sizes have to be tuned.

\subsection{Divergence}
As threads are executed by warps on the GPU SIMD units, applications should have
regular patterns for memory accesses and control flow.

The regularity of memory accesses patterns is important to achieve high memory
throughput. As the threads within a warp load data from memory concurrently, the
developer has to coalesce accesses to device memory and avoid bank conflicts in
the shared memory \cite[chap. 6]{nvidia_best_practice}. This can be done by
reorganizing data storage.

\alert{The} regularity of control flow is important to achieve high instruction
throughput, and is obtained when all the threads within a warp execute the same
instruction concurrently \cite[chap.
9]{nvidia_best_practice}. 
In fact, when the threads of a same warp diverge (i.e. they follow different
execution paths), the different execution paths are serialized.  For an {\it if}
statement, the {\it then} and {\it else} branches are serially executed. For a
loop, any thread exiting the loop has to wait until all the threads of its warp
exit the loop.  In the following we will distinguish branch divergence due to
{\it if} statements and loop divergence due to loop statements.

The impact of branch divergence can be statically estimated by counting the
number of instructions issued within the scope of the {\it if} statement. Let us
consider the {\it then} branch issues $n_{then}$ instructions and the {\it else}
branch issues $n_{else}$ instructions. If the warp does not diverge, either
$n_{then}$ or $n_{else}$ instructions are issued depending on the evaluation of
the condition. If the warp diverges, $n_{then} + n_{else}$ instructions are
issued.

Contrary to branch divergence, measuring the impact of loop divergence requires
a dedicated indicator and profiling. We introduced in \cite{Gouicem12} the mean
deviation to the maximum of a warp. This indicator is similar to the standard
deviation, which is the mean deviation to the mean value. However, as the number
of loop iterations issued for a warp is equal to the maximum number of loop
iterations issued by any thread within the warp, it is relevant to consider the
mean deviation to the maximum value. This gives the mean number of loop
iterations a thread remains idle within its warp.
More formally, we denote $\ell_i$ the number of loop iterations of the thread
$i$ and we number the threads within a warp from $1$ to $32$. If $\ell =
\{\ell_i, i\in \llbracket 1, 32 \rrbracket\}$, the Mean Deviation to the Maximum
(MDM) of a warp is defined as
$$\text{MDM}({\ell}) = \text{max}(\ell) - \text{mean}(\ell).$$

\noindent We can normalize the mean deviation to the maximum by
$\displaystyle{\text{max}(\ell)}$ to compute the average percentage of loop
iterations for which a thread remains idle within its warp. Hence, the
Normalized Mean Deviation to the Maximum (NMDM) is
$$\text{NMDM}(\ell) = 1 - \frac{\text{mean}(\ell)}{\text{max}(\ell)}.$$



\section{Mathematical preliminaries}
\label{sect_math_pre}

In this section we give some definitions to introduce more formally the Table
Maker's Dilemma.  We also recall some known properties on the distribution of
the elements of the set \mbox{$\{a\cdot x \mod{1}~|~x<n \}$} with a fixed
\cite{Slater67}, as well as the corresponding continued fraction formalism.

\subsection{The Table Maker's Dilemma}
\label{sect:tmd}

Before defining the Table Maker's Dilemma, we introduce some notations and
definitions. We denote $\{X\}$ or $X\mod{1}$ the fractional part of $X$. We
write $X \cmod{1}$ the centered modulo, which is the real $Y$ such that $X-Y \in
\mathbb{Z}$ and $Y\in~]{-1/2}, 1/2]$ ($Y$ equals $X - \lfloor X \rfloor$ or $X -
\lceil X \rceil$ depending on which has the lowest absolute value). \alert{We
  also write $\mathbb{F}_p$ the set of precision-$p$ floating point numbers and
  $\#_pE$ the number of precision-$p$ floating-point numbers in the set $E$
  (namely $\#(E \cap \mathbb{F}_p)$\;)}.

\begin{definition}
  The mantissa $m(x)$ and \alert{the} exponent $e(x)$ of a non-zero real number
  $x$ are defined by $|x| = m(x)\cdot 2^{e(x)}$ with $1/2 \leq m(x) < 1$.
\end{definition}

\begin{definition}
  We define $\dist{p}{x} = | 2^p \cdot m(x) \cmod{1} |$ as the \alert{scaled}
  distance between a real number $x$ and the closest precision-$p$
  floating-point number.
  \label{def:dist}
\end{definition}

\begin{definition}
  We now define a \emph{$(p,\epsilon)$ hard-to-round case} (or \emph{HR-case})
  of a real-valued function $f$ as a precision-$p$ floating-point number $x$
  solution of the inequality $$\dist{p}{f(x)} < \epsilon.$$
\label{eq_hr_cas}
\end{definition}

\alert{The given definition of HR-case only applies for directed
  rounding. However, this definition can be extended to all IEEE-754 rounding
  modes as rounding-to-nearest $(p,\epsilon)$ HR-cases are directed rounding
  $(p+1, 2\epsilon)$ HR-cases. To simplify notations, we will then focus on
  directed rounding HR-cases.}

It has to be noticed that if $x$ is a $(p,\epsilon)$ hard-to-round case, it also
satisfies $ 2^p \cdot m(f(x)) + \epsilon < 2\epsilon \mod{1}.$ \alert{The latter
  inequality is used to test if an argument is a $(p,\epsilon)$ HR-case as it
  avoids the computation of absolute values and $\text{cmod}$}.

Hence, a $(p,2^{-p'})$ HR-case $x$ is a precision-$p$ floating-point number for
which $f(x)$ is at a scaled distance (as defined in Def. \ref{def:dist}) less
than $2^{-p'}$ from the closest precision-$p$ floating-point number. In other
words, more than $p+p'$ bits of accuracy are necessary to correctly round $f(x)$
at precision-$p$.

\begin{definition}[Table Maker's Dilemma]
  If $f$ is a real valued function defined over a domain $D$, we define the
  Table Maker's Dilemma as finding a non trivial lower bound on
  $\{\dist{p}{f(x)}, x\in D\}$.
\label{def_tmd}
\end{definition}

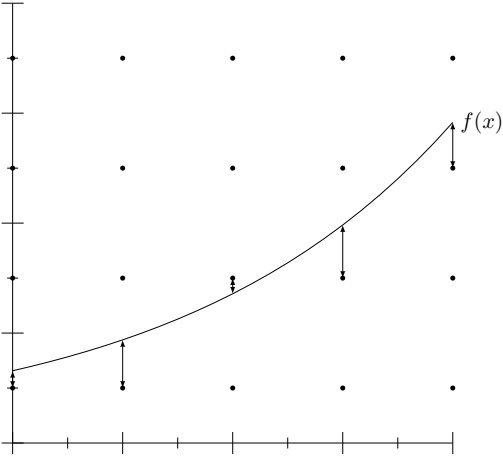
\begin{figure}
  \centering
  \resizebox{0.8\linewidth}{!}{

  \begin{tikzpicture}[>=latex] 
    \pgfmathsetmacro\intervalSize{7} 
    \pgfmathsetmacro\nbpts{4} 
    \pgfmathsetmacro\tangente{0.1} 
    \pgfmathsetmacro\CoeffEx{1.6} 
    \pgfmathsetmacro\AddEx{0.15} 
    \pgfmathsetmacro\eps{0.3}
    
    \pgfmathsetmacro\bp{0} 
    \pgfmathsetmacro\ep{\intervalSize} 
    \pgfmathsetmacro\nbptsBis{\nbpts-1} 
    \pgfmathsetmacro\step{(\ep-\bp)/\nbpts}  
    \pgfmathsetmacro\ptsH{ \tangente*\step} 
    \pgfmathsetmacro\midptsH{\ptsH/2} 
    \pgfmathsetmacro\shift{(((\ep-\bp)/2)*\ptsH)/\step}  

    \draw (\bp,0) -- (\ep,0);
    \draw (0,\bp) -- (0,\ep);

    \foreach \x in {0,...,\nbpts}{ 
      \draw (\bp + \x*\step, -\ptsH) 
      -- (\bp + \x*\step, \ptsH); 
      \draw (-\ptsH, \bp + \x*\step) 
      -- (\ptsH, \bp + \x*\step); 
    }
    
    \foreach \x in {0,...,\nbptsBis}{ 
      \draw (\bp + \x*\step + 0.5*\step,-\midptsH) 
      -- (\bp + \x*\step + 0.5*\step, \midptsH); 
      \draw (-\midptsH, \bp + \x*\step + 0.5*\step) 
      -- (\midptsH, \bp + \x*\step + 0.5*\step); 
    }

    \foreach \x in {0,...,\nbpts}{ 
      \foreach \y in {0,...,\nbptsBis}{ 
        \fill (\bp + \x*\step,\bp + \y*\step + 0.5*\step) circle (0.04);
      }
    }
    \draw[domain=\bp:\ep] plot (\x, { exp(\CoeffEx*\x/\intervalSize) + \AddEx })node[anchor=west] {$f(x)$};

    \foreach \x in {0,...,\nbpts}{ 
      \pgfmathsetmacro\Ex{exp(\CoeffEx*(\bp + \x*\step)/\intervalSize) + \AddEx}
      \pgfmathsetmacro\FloorEx{(floor((\Ex- 0.5*\step)/\step) + 0.5)*\step}
      \pgfmathsetmacro\CeilEx{(ceil((\Ex- 0.5*\step)/\step)+0.5)*\step}
      \pgfmathsetmacro\RoundEx{
        ifthenelse(
        less(abs(\FloorEx-\Ex), abs(\CeilEx-\Ex)),
        \FloorEx,
        \CeilEx)}

      \draw[ultra thin, <->] 
      (\bp + \x*\step, \RoundEx) 
      -- ( \bp + \x*\step, \Ex);
    }
  \end{tikzpicture}
  }
  \caption{Distances between the curve defined by $f$ and the rounding
    breakpoints for rounding-to-nearest.}
  \label{fig_dist_curve_float}
\end{figure}

We call {\it hardest-to-round} cases the arguments $x\in D$ minimizing
$\dist{p}{f(x)}$. Knowing the hardest-to-round cases gives us a lower bound on
the distances between the function $f$ and the rounding breakpoints (see Fig.
\ref{fig_dist_curve_float}) and therefore a solution to the TMD.

\medskip

The general method to find the hardest-to-round cases of a function is the
following:
\begin{enumerate}
\item fix a ``convenient'' $\epsilon$ using probabilistic assumptions
  \cite[Sect.  12.2]{hfpa},
\item find $(p, \epsilon)$ HR-cases with \emph{ad hoc} methods such as Lefèvre
  or SLZ algorithms,
\item find the hardest-to-round among the $(p, \epsilon)$ HR-cases using Ziv
  method \cite{Ziv91}.
\end{enumerate}

The most compute intensive step in this method is the second one. Lefèvre or SLZ
algorithms both relies on the following \alert{three} major steps.
\begin{enumerate}
\item \alert{{\it The split of the domain of definition of the function }: we
    split the domain of definition of the function in $d$ domains $D_i = [X_i,
    X_{i+1} [~\cap~\mathbb{F}_p$ such that, $e(x)=e(y)~\forall x,y \in D_i$ and
    $\#_pD_i = \#_pD_j~\forall i, j \in \llbracket 0, d-1 \rrbracket$}.
\item {\it \alert{The} generation of polynomial
    approximations \label{pol_approx_step}}: given a relative error
  $\epsilon_{approx}$, we approximate the function $f(X)$ with \mbox{$X\in D_i$}
  by polynomials $P_i(x)$ with \mbox{$x\in \llbracket 0, \#_p D_i - 1
    \rrbracket$} such that \mbox{$|P_i(x) - f(X)| <
    \epsilon_{approx}2^{e(f(X))-p}$}. \alert{We thus proceed to a change of
    variable enabling to test the floating-point arguments $X\in D_i$ by testing
    the integers \mbox{$x\in \llbracket 0, \#_p D_i - 1 \rrbracket$}.  Each
    polynomial $P_i$ is \alert{first centered on the domain} $D_i$ by applying
    the change of variable \mbox{$\phi_1 : X \mapsto X - X_i$}. Then, as the
    exponent is constant over each domain $D_i$, we will consider integer
    arguments by applying the change of variable \mbox{$\phi_2 : X \mapsto
      X\cdot 2^{p-e(X_i)}$}. All in all, 
    \mbox{$x = \phi_2 \circ \phi_1 (X) = 2^{p-e(X_i)}(X-X_i)$}}.
\item {\it \alert{The} HR-case search}: we find the $(p, \epsilon')$ HR-cases of
  $P_i$ with $\epsilon' = \epsilon+\epsilon_{approx}$ which are the
  $(p,\epsilon)$ HR-cases for $f$ in $D_i$.
\end{enumerate}

In the HR-case search of both algorithms, a Boolean test is used to isolate
HR-cases. It successes if there is no $(p, \epsilon')$ HR-case
for $P_i$ in $D_i$ and fails otherwise.


In this paper, we focus on Lefèvre algorithm which truncates polynomials $P_i$
to degree one for the Boolean test. \alert{We denote $Q_i(x) = P_i(x) \mod{x^2}$
  the truncation of $P_i$ to degree one with \mbox{$|Q_i(x) - P_i(x)| <
    \epsilon_{trunc}2^{e(P_i(x))-p}$}}, and
$$2^p\cdot m(Q_i(x)) + \epsilon''= b - a\cdot x,$$ 
with $\epsilon'' = \epsilon' + \epsilon_{trunc}$. Hence, the Boolean test of
Lefèvre algorithm consists of testing if the following inequality holds:
\begin{equation}
\label{eq_inf}
\min{\{b-a\cdot x \mod{1}~|~x<\#_pD_i \}} < 2\epsilon''.
\end{equation}
More precisely, if the inequality \eqref{eq_inf} does not hold, \alert{the
  Boolean test returns Success as} there is no $(p, \epsilon'')$ HR-cases for
$Q_i$ in $D_i$ which implies there is no $(p, \epsilon')$ HR-case for $P_i$ in
$D_i$. \alert{Else, it returns Failure}. Moreover, we remark that computing the
minimum of the set \mbox{$\{b-a\cdot x \mod{1}~|~x < \#_pD_i \}$} is similar to
finding the multiple of $a$ which is the closest to the left of $b$ modulo $1$
on the unit segment.




\newcounter{cpt}

\subsection{Properties of the set \mbox{$\{a\cdot x \mod{1}~|~x<n \}$}}
\label{section_three_length_theorem}

Here we will detail some properties on the configurations of the points
\mbox{$\{a\cdot x \mod{1}~|~x<n \}$} over the unit segment. These properties are
necessary to efficiently locate the closest point to $\{b\}$ in these
configurations \alert{as we need a Boolean test over a lower bound on
  \mbox{$\{b-a\cdot x \mod{1}~|~x < \#_pD_i \}$}}.

\begin{theorem}[Three distance theorem \cite{Slater50}]
  \label{theorem_three_distance}
  Let \mbox{$0 < a < 1$} be an irrational number. If we place on the unit
  segment $[0, 1[$ the points $\{0\}$, $\{a\}$, $\{2a\}$, $\dots$, $\{(n-1)a\}$,
  these points partition the unit segment into $n$ intervals having at most
  three lengths with one being the sum of the two others.
\end{theorem}

\begin{figure}[!h]
  \centering
  \resizebox{\linewidth}{!}{

  \begin{tikzpicture}
    \pgfmathsetmacro\a{14} 
    \pgfmathtruncatemacro\modulus{45} 

    \pgfmathsetmacro\ptsH{0.25} 
    \pgfmathsetmacro\whiteSpace{1.5} 
    \pgfmathsetmacro\lineScale{0.3} 
    \pgfmathsetmacro\blPos{\modulus + 0.05*\modulus} 
    \pgfmathsetmacro\nLabelPos{-1}

    \foreach \y in {0,...,15}{ 
      \draw (0, -\whiteSpace*\y) -- (\lineScale*\modulus, -\whiteSpace*\y); 
      \draw (\lineScale*0, -\whiteSpace*\y+\ptsH) -- (\lineScale*0, -\whiteSpace*\y-\ptsH) node [at end, below] {$0$};
      \draw (\lineScale*\modulus, -\whiteSpace*\y+\ptsH) -- (\lineScale*\modulus, -\whiteSpace*\y-\ptsH);
      \foreach \x in {0,...,\y}{
        \pgfmathsetmacro\Pts{Mod(\x*\a,\modulus)}
        \draw (\lineScale*\Pts, -\whiteSpace*\y+\ptsH) -- (\lineScale*\Pts, -\whiteSpace*\y-\ptsH) node [at end, below] {$\x$} node [midway, above] {};
      }
    }
    
    \pgfmathsetmacro\N{0}
    \node [anchor=south] at (\lineScale*22.5, -\whiteSpace*\N) {\small \modulus};
    \node at (\lineScale*\blPos, -\whiteSpace*\N) {$\mathbf{(0, 0)}$};
    \pgfmathsetmacro\N{1}
    \setcounter{cpt}{0}
    \foreach \x in {14, 31}{
      \pgfmathsetmacro\pt{\value{cpt}+(\x*0.5)}
      \node [anchor=south] at (\lineScale*\pt, -\whiteSpace*\N) {\small $\x$};
      \addtocounter{cpt}{\x}
    }
    \node at (\lineScale*\blPos, -\whiteSpace*\N) {$\mathbf{(0, 1)}$};

    \pgfmathsetmacro\N{2}
    \setcounter{cpt}{0}
    \foreach \x in {14, 14, 17}{
      \pgfmathsetmacro\pt{\value{cpt}+(\x*0.5)}
      \node [anchor=south] at (\lineScale*\pt, -\whiteSpace*\N) {\small $\x$};
      \addtocounter{cpt}{\x}
    }
    \node at (\lineScale*\blPos, -\whiteSpace*\N) {$\mathbf{(0, 2)}$};

    \pgfmathsetmacro\N{3}
    \setcounter{cpt}{0}
    \foreach \x in {14, 14, 14, 3}{
      \pgfmathsetmacro\pt{\value{cpt}+(\x*0.5)}
      \node [anchor=south] at (\lineScale*\pt, -\whiteSpace*\N) {\small $\x$};
      \addtocounter{cpt}{\x}
    }
    \node at (\lineScale*\blPos, -\whiteSpace*\N) {$\mathbf{(1, 0)}$};

    \pgfmathsetmacro\N{4}
    \setcounter{cpt}{0}
    \foreach \x in {11, 3, 14, 14, 3}{
      \pgfmathsetmacro\pt{\value{cpt}+(\x*0.5)}
      \node [anchor=south] at (\lineScale*\pt, -\whiteSpace*\N) {\small $\x$};
      \addtocounter{cpt}{\x}
    }

    \pgfmathsetmacro\N{5}
    \setcounter{cpt}{0}
    \foreach \x in {11, 3, 11, 3, 14, 3}{
      \pgfmathsetmacro\pt{\value{cpt}+(\x*0.5)}
      \node [anchor=south] at (\lineScale*\pt, -\whiteSpace*\N) {\small $\x$};
      \addtocounter{cpt}{\x}
    }

    \pgfmathsetmacro\N{6}
    \setcounter{cpt}{0}
    \foreach \x in {11, 3, 11, 3, 11, 3, 3}{
      \pgfmathsetmacro\pt{\value{cpt}+(\x*0.5)}
      \node [anchor=south] at (\lineScale*\pt, -\whiteSpace*\N) {\small $\x$};
      \addtocounter{cpt}{\x}
    }
    \node at (\lineScale*\blPos, -\whiteSpace*\N) {$\mathbf{(1, 1)}$};

    \pgfmathsetmacro\N{7}
    \setcounter{cpt}{0}
    \foreach \x in {8, 3, 3, 11, 3, 11, 3, 3}{
      \pgfmathsetmacro\pt{\value{cpt}+(\x*0.5)}
      \node [anchor=south] at (\lineScale*\pt, -\whiteSpace*\N) {\small $\x$};
      \addtocounter{cpt}{\x}
    }

    \pgfmathsetmacro\N{8}
    \setcounter{cpt}{0}
    \foreach \x in {8, 3, 3, 8, 3, 3, 11, 3, 3}{
      \pgfmathsetmacro\pt{\value{cpt}+(\x*0.5)}
      \node [anchor=south] at (\lineScale*\pt, -\whiteSpace*\N) {\small $\x$};
      \addtocounter{cpt}{\x}
    }

    \pgfmathsetmacro\N{9}
    \setcounter{cpt}{0}
    \foreach \x in {8, 3, 3, 8, 3, 3, 8, 3, 3, 3}{
      \pgfmathsetmacro\pt{\value{cpt}+(\x*0.5)}
      \node [anchor=south] at (\lineScale*\pt, -\whiteSpace*\N) {\small $\x$};
      \addtocounter{cpt}{\x}
    }
    \node at (\lineScale*\blPos, -\whiteSpace*\N) {$\mathbf{(1, 2)}$};

    \pgfmathsetmacro\N{10}
    \setcounter{cpt}{0}
    \foreach \x in {5, 3, 3, 3, 8, 3, 3, 8, 3, 3, 3}{
      \pgfmathsetmacro\pt{\value{cpt}+(\x*0.5)}
      \node [anchor=south] at (\lineScale*\pt, -\whiteSpace*\N) {\small $\x$};
      \addtocounter{cpt}{\x}
    }

    \pgfmathsetmacro\N{11}
    \setcounter{cpt}{0}
    \foreach \x in {5, 3, 3, 3, 5, 3, 3, 3, 8, 3, 3, 3}{
      \pgfmathsetmacro\pt{\value{cpt}+(\x*0.5)}
      \node [anchor=south] at (\lineScale*\pt, -\whiteSpace*\N) {\small $\x$};
      \addtocounter{cpt}{\x}
    }

    \pgfmathsetmacro\N{12}
    \setcounter{cpt}{0}
    \foreach \x in {5, 3, 3, 3, 5, 3, 3, 3, 5, 3, 3, 3, 3}{
      \pgfmathsetmacro\pt{\value{cpt}+(\x*0.5)}
      \node [anchor=south] at (\lineScale*\pt, -\whiteSpace*\N) {\small $\x$};
      \addtocounter{cpt}{\x}
    }
    \node at (\lineScale*\blPos, -\whiteSpace*\N) {$\mathbf{(1, 3)}$};

    \pgfmathsetmacro\N{13}
    \setcounter{cpt}{0}
    \foreach \x in {2, 3, 3, 3, 3, 5, 3, 3, 3, 5, 3, 3, 3, 3}{
      \pgfmathsetmacro\pt{\value{cpt}+(\x*0.5)}
      \node [anchor=south] at (\lineScale*\pt, -\whiteSpace*\N) {\small $\x$};
      \addtocounter{cpt}{\x}
    }

    \pgfmathsetmacro\N{14}
    \setcounter{cpt}{0}
    \foreach \x in {2, 3, 3, 3, 3, 2, 3, 3, 3, 3, 5, 3, 3, 3, 3}{
      \pgfmathsetmacro\pt{\value{cpt}+(\x*0.5)}
      \node [anchor=south] at (\lineScale*\pt, -\whiteSpace*\N) {\small $\x$};
      \addtocounter{cpt}{\x}
    }

    \pgfmathsetmacro\N{15}
    \setcounter{cpt}{0}
    \foreach \x in {2, 3, 3, 3, 3, 2, 3, 3, 3, 3, 2, 3, 3, 3, 3, 3}{
      \pgfmathsetmacro\pt{\value{cpt}+(\x*0.5)}
      \node [anchor=south] at (\lineScale*\pt, -\whiteSpace*\N) {\small $\x$};
      \addtocounter{cpt}{\x}
    }
    \node at (\lineScale*\blPos, -\whiteSpace*\N) {$\mathbf{(2, 0)}$};
    
  \end{tikzpicture}
  }
  \caption{Example of configurations generated by \mbox{$a = 14/45$}. The unit
    segment is scaled by a factor $45$ for clarity. Each two-length
    configuration is labelled by its index $(i,t)$ on the right.}
  \label{fig:three_length}
\end{figure}
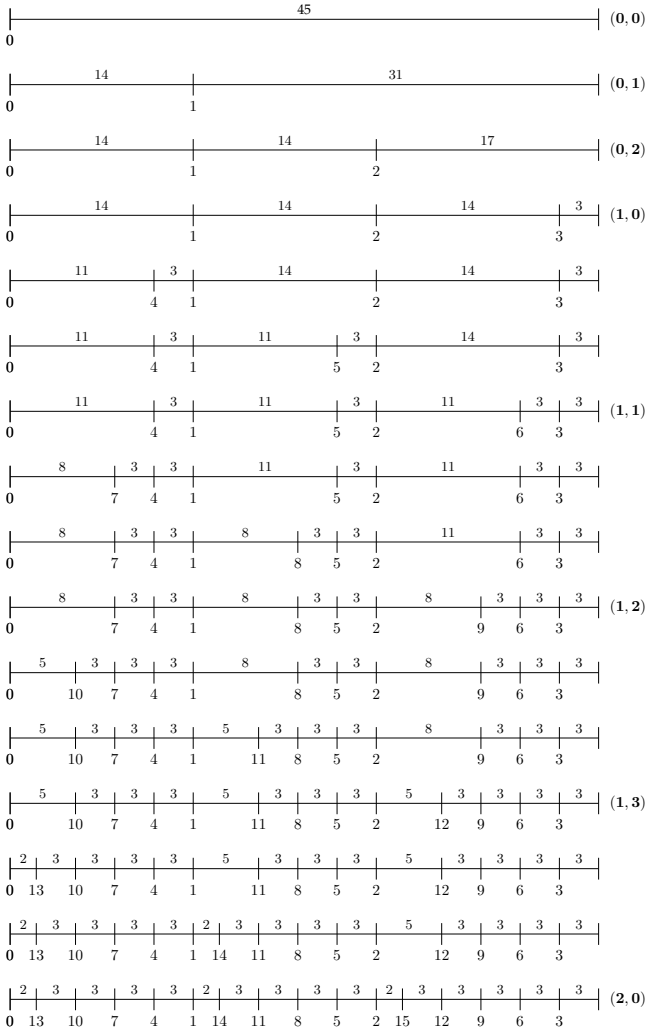

\alert{ Actually, the lengths and the distribution of these lengths heavily rely
  on the continued fraction expansion of $a$ \cite{Slater67}, which we will
  denote by $$ a = \cfrac{1}{ k_1 + \cfrac{1}{ k_2 + \ddots}}.$$}

\alert{ We denote by $(\theta_i)_{i\in\mathbb{N}}$ the sequence of the
  remainders computed during the continued fraction expansion of $a$ using the
  Euclidean algorithm, and by $(p_i/q_i)_{i\in\mathbb{N}}$ the sequence of the
  convergents of $a$, defined by the following recurrence relations :
\[
\begin{array}{lll}
  p_{-1} = 1 & p_0 = 0 & p_{i+1}= p_{i-1} + k_{i+1}\cdot p_{i}, \\
  q_{-1} = 0 & q_0 = 1 & q_{i+1}= q_{i-1} + k_{i+1}\cdot q_{i}, \\
  \theta_{-1} = 1 & \theta_0 = a & \theta_{i+1} = \theta_{i-1} - k_{i+1}\cdot\theta_{i},
\end{array}
\]
with $k_{i+1} = \lfloor \theta_{i-1}/\theta_i \rfloor$.  It has to be noticed
that $(\theta_i)_{i\in\mathbb{N}}$ is a decreasing real-valued positive sequence
whereas $(p_i)_{i\in\mathbb{N}}$ and $(q_i)_{i\in\mathbb{N}}$ are increasing
integer-valued positive sequences.  We also define \mbox{$\theta_{i-1,t} =
  \theta_{i-1} - t\cdot\theta_{i}$} and \mbox{$q_{i-1, t} = q_{i-1} + t\cdot
  q_i$} with \mbox{$t\in \llbracket 0, k_{i+1}\llbracket$}.  The lengths
obtained when adding multiples of $a$ over the unit segment are therefore the
elements of the sequence 
$(\theta_{i,t})_{i\in\mathbb{N},t \in \llbracket 0, k_{i+1}\llbracket}$
 \cite{Slater67}}. An example is provided in Fig. \ref{fig:three_length} and
Table \ref{tab:detailed_example}. \alert{ All the properties provided in this
  section are valid when $a$ is irrational. However, they are also valid for $a$
  rational as long as $\theta_i \not= 0$ (that is to say, until the last
  quotient of the continued fraction expansion is computed)}.

\begin{table}[t]
  \centering
  \begin{tabular}{cccccc|c}
    \hline
    i & t & $q_{i-1,t}$ & $q_i$ & $\theta_{i-1, t}$ & $\theta_i$ & $k_{+1}i$ \\
    \hline
    \multirow{3}{*}{0} & 0 & 0 & \multirow{3}{*}{1} & 45 & \multirow{3}{*}{14} & \multirow{3}{*}{3}\\
    & 1 & 1 &  & 31 & &\\
    & 2 & 2 &  & 17 & &\\
    \hline
    \multirow{4}{*}{1} & 0 & 1 & \multirow{4}{*}{3} & 14 & \multirow{4}{*}{3} & \multirow{4}{*}{4}\\
    & 1 &  4 & & 11 & &\\
    & 2 &  7 & &  8 & &\\
    & 3 & 10 & &  5 & &\\
    \hline
    2 & 0 & 3 & 13 & 3 & 2 & 1\\
    \hline
    \multirow{2}{*}{3} & 0 & 13 & \multirow{2}{*}{16} & 2 & \multirow{2}{*}{1} & \multirow{2}{*}{2}\\
    & 1 & 29 & & 1 && \\
    \hline
    4 & 0 & 16 & 45 & 1 & 0 & 1\\
    \hline
  \end{tabular}
  \caption{Values of $\theta_i$, $\theta_{i-1,t}$, $q_i$, $q_{i-1,t}$ and $k_i$ for each two-length configuration of the example of Fig. 3. As in Fig. 3, the lengths $\theta_i$ and $\theta_{i-1, t}$ are scaled by a factor 45.}
  \label{tab:detailed_example}
\end{table}

\alert{ In the following, we will use some properties on the configurations
  \mbox{$\{a\cdot x \mod{1}~|~x<n \}$} which contain intervals of only two
  different lengths. They are of algorithmic interest as there are only $O(\log
  n)$ such configurations when $n$ tends to infinity. 
   Each label $(i,t)$,  with $i \in
  \mathbb{N}$ and $t \in \llbracket 0, k_{i+1}\llbracket$, 
   denotes 
   one two-length configuration which verifies the following
  equation
\begin{equation}
  \label{eq:config_sous}
  q_i\cdot\theta_{i-1, t} + q_{i-1,t}\cdot\theta_i = 1.
\end{equation}
This Equation \eqref{eq:config_sous} gives details on the number of intervals of
each length. After adding $q_i + q_{i-1, t}$ multiples of $a \mod{1}$ over the
unit segment, there are exactly two different lengths of intervals over the unit
segment : $q_i$ intervals of length $\theta_{i-1, t}$ and $q_{i-1, t}$ intervals
of length $\theta_i$}.

\alert{A special and noticeable subset of the two-length configurations
  corresponds to the configurations produced using the division-based Euclidean
  algorithm. These are the $(i, 0)$ configurations, satisfying
\begin{equation}
  \label{eq:config_div}
  q_i\cdot\theta_{i-1} + q_{i-1}\cdot\theta_i = 1.
\end{equation}
Furthermore we have a way to construct the two-length configurations.
\begin{property}[Two-length configurations construction \cite{Slater67}]
  \label{prop:2length_const}
  Given a two-length configuration $(i, t)$ for some $i\in\mathbb{N}$ and $0 \le
  t < k_{i+1}$, the next two-length configuration is 
  \[\left\{\begin{array}{ll}
      (i, t+1)& \text{if}~t < k_{i+1}-1,\\
      (i+1, 0)& \text{if}~t = k_{i+1}-1.\\
    \end{array}\right.
  \]
\end{property}}

\alert{ To simplify notations, given the $(i,t)$ configuration we will write
  $(i,t+1)$ its next two-length configuration and assimilate the configuration
  $(i, k_{i+1})$ to $(i+1, 0)$. Property \ref{prop:2length_const} implies that
  for going from a two-length configuration to the next, the intervals of length
  $\theta_{i-1, t}$ are split. The way intervals are split is described by the
  following \emph{directed reduction} property and illustrated in
  Fig. \ref{fig:three_length}}.


\alert{
\begin{property}[Directed reduction \cite{Ravenstein88}]
  \label{prop:directed_reduction}
  Given the two-length configuration $(i, t)$, when constructing the next
  two-length configuration, intervals of length $\theta_{i-1, t}$ are split
  into two intervals in this order from left to right :
  \begin{itemize}
  \item one of length $\theta_i$ and one of length $\theta_{i-1, t+1}$ if $i$ is
    even,
  \item one of length $\theta_{i-1, t+1}$ and one of length $\theta_i$ if $i$ is
    odd.
  \end{itemize}
\end{property}}


\section{HR-case search on GPU}
\label{sect_hr_search}

In this section we describe two algorithms for HR-case search: Lefèvre HR-case
search \alert{originally described in \cite{Lefevre2005} and our new and more
  regular HR-case search. Both algorithms make use of Boolean tests which rely
  on the properties described in
  Sect. \ref{section_three_length_theorem}. Hence, we will describe both of them
  with continued fraction expansions which give a uniform formalism to explain
  and compare their different behaviours. Then we will describe how they have
  been deployed on GPU and the benefit on divergence provided by our new
  algorithm}.

\subsection{Lefèvre HR-case search}
\label{sect_lefevre_algo}

In \cite{Lefevre2005}, Lefèvre presented an algorithm to search for $(p,
\epsilon')$ HR-cases of a polynomial $P_i(x)$. This algorithm relies on a
Boolean test on $Q_i(x)$ (the truncation of $P_i(x)$ to degree one) which
computes a lower bound of the set \mbox{$\{b - a\cdot x \mod{1}~|~x<\#_pD_i\}$}
and \alert{returns Success} if the inequality \eqref{eq_inf} \alert{does not}
hold, \alert{Failure otherwise}.

In Sect. \ref{section_three_length_theorem}, we described some properties of the
configurations \mbox{$\{a\cdot x \mod{1}~|~x<n \}$}. According to these
properties, computing the lengths of the intervals of the two-length
configurations can be done efficiently in $O(\log{\#_pD_i})$ arithmetic
operations by computing the continued fraction expansion of $a$. However, if we
use continued fraction expansion, we will place more points than $\#_pD_i$ on
the unit segment (at most $2 \cdot \#_pD_i$ if we use the subtraction-based
Euclidean algorithm). To take advantage of the efficient construction of the
two-length configurations, Lefèvre HR-case search computes the minimum of
\mbox{$\{b - a\cdot x \mod{1}~|~x<n\}$} with $n$ the number of multiples of $a$
placed and $n \ge \#_pD_i$\,. This gives a lower bound on \mbox{$\{b - a\cdot x
  \mod{1}~|~x<\#_pD_i\}$}. Then the minimum of \mbox{$\{\dist{p}{P_i(x)} <
  \epsilon'~|~x<\#_pD_i\}$} is exactly computed by exhaustive search in
$O(\#_pD_i)$ arithmetic operations only if required. To minimize this exhaustive
search we use a filtering strategy in three phases.
\begin{itemize}[\IEEEsetlabelwidth{}]
\item Phase 1: we compute a lower bound on \mbox{$\{b-a\cdot x
    \mod{1}~|~x<\#_pD_i\}$} and test if this lower bound matches a $(p,
  \epsilon'')$ HR-case of $Q_i$. If not there is no $(p,\epsilon')$ HR-case for
  $P_i$ in $D_i$. Else, go to next phase.
\item Phase 2: we split $D_i$ in \alert{sub-domains} $D_{i,j}$, we refine the
  approximation $Q_i(x)$ by $Q_{i,j}(x)$ and we compute a lower bound on
  \mbox{$\{b_j-a_j\cdot x \mod{1}~|~x<\#_pD_{i,j}\}$} for each $D_{i,j}$. For
  each $D_{i,j}$ where the lower bound on \mbox{$\{b_j-a_j\cdot x
    \mod{1}~|~x<\#_pD_{i,j}\}$} matches a $(p, \epsilon''_j)$ HR-case of
  $Q_{i,j}$, go to next phase.
\item Phase 3: we search exhaustively for $(p, \epsilon')$ of $P_i$ in $D_{i,j}$
  using the table difference method (see Sect. \ref{sect_pol_approx}).
\end{itemize}

\medskip 

The corner stone of Lefèvre algorithm strategy is therefore the computation of
the minimum of \mbox{$\{b-a\cdot x \mod{1}~|~x<n\}$}. In other words, it
computes the distance between $\{b\}$ and the closest point to the left of
$\{b\}$ in the configuration \mbox{$\{a\cdot x \mod{1}~|~x<n \}$}. We write $N$
the number of floating-point numbers in the considered sub-domain ($n\ge N$ as
we compute a lower bound). Depending on how we generate the two-length
configurations (using the subtraction-based or the division-based Euclidean
algorithm) we can derive from Property \ref{prop:directed_reduction} two ways to
compute this distance. The first one is Lefèvre HR-case search, the second one
is the new HR-case search proposed in Sect. \ref{sect_new_algo}.


\begin{algorithm}[t]
  \footnotesize
  \SetKwInOut{Input}{input}
  \SetKwInOut{Output}{output}
  \Input{$b - a\cdot x$, $\epsilon''$, $N$}
  {\bf initialisation: }
  \begin{tabular}{lll}
    $p \leftarrow \{a\}$; & $q \leftarrow 1 - \{a\}$; & $d \leftarrow \{b\}$; \\
    $u \leftarrow 1$; & $v \leftarrow 1$; &
  \end{tabular}

  \lIf{ $d < \epsilon''$ }{\Return Failure\;}
  \While {True}
  {
    \eIf{$d < p$}
    {
      $k = \lfloor q/p \rfloor$\;
      $q \leftarrow q - k*p$; $u \leftarrow u+ k*v$\;
      \lIf{$u+v \ge N$}{\Return Success\;}
      $p \leftarrow p - q$; $v \leftarrow v + u$\;
    }
    {
      $d \leftarrow d - p$\;
      \lIf{$d < \epsilon''$}{\Return Failure\;}
      $k = \lfloor p/q \rfloor$\;
      $p \leftarrow p - k*q$; $v \leftarrow v+ k*u$\;
      \lIf{$u+v \ge N$}{\Return Success\;}
      $q \leftarrow q - p$; $u \leftarrow u + v$\;
    }
  }
  \caption{Lefèvre lower bound computation and test algorithm.}
  \label{algo_lefevre}
\end{algorithm}

In the lower bound computation of Lefèvre HR-case search, the way the two-length
configurations are computed depends on the length of the interval containing
$\{b\}$. When adding points in the interval containing $\{b\}$ and in the
direction of $\{b\}$ Lefèvre uses a subtraction-based Euclidean algorithm (he
moves from the $(i,t)$ configuration to $(i, t+1)$). Otherwise he uses a
division-based Euclidean algorithm (he moves from the $(i,t)$ configuration to
$(i+1, 0)$).  Algorithm \ref{algo_lefevre} describes the lower bound computation
of Lefèvre HR-case search and the corresponding test with respect to
$\epsilon''$ which is the sum of all errors involved.

\alert{In this algorithm, the variables $u$ and $v$ count the number of
  intervals as in Equation \eqref{eq:config_sous} in order to exit when $n = u+v
  \ge N$ : $u$ and $v$ store respectively $q_i$ and $q_{i-1, t}$ for $i$ even
  and $q_{i-1, t}$ and $q_i$ for $i$ odd. The variables $p$ and $q$ store
  respectively the lengths $\theta_i$ and $\theta_{i-1,t}$ for $i$ even, and the
  lengths $\theta_{i-1, t}$ and $\theta_i$ for $i$ odd.  The variable $d$
  contains the distance between $\{b\}$ and the closest multiple $\{a\cdot x\}$
  to its left}.

\begin{figure*}
  \centering
  \subfloat[$i$ is even]{
    \centering
    \resizebox{0.45\linewidth}{!}{
      \begin{tikzpicture}[>=latex]
  \pgfmathsetmacro\a{6} 
  \pgfmathtruncatemacro\modulus{45} 
  
  \pgfmathsetmacro\ptsH{0.25} 
  \pgfmathsetmacro\lineScale{0.3} 

  \draw (0, 0) -- (\lineScale*\modulus, 0);
  \draw (\lineScale*0, \ptsH) -- (\lineScale*0, -\ptsH) 
  ;
  \draw (\lineScale*\modulus, \ptsH) -- (\lineScale*\modulus, -\ptsH) 
  ;

  \fill (\lineScale*20, 0) circle (\lineScale*0.25) node [below] {$b$};
  
  \foreach \x in {1,...,4}{
    \pgfmathsetmacro\Pts{Mod(\x*\a,\modulus)}
    \draw (\lineScale*\Pts, \ptsH) -- (\lineScale*\Pts, -\ptsH) node [at end, below] {$(i, \x)$};
  }
  \foreach \x in {5,...,6}{
    \pgfmathsetmacro\Pts{Mod(\x*\a,\modulus)}
    \draw [densely dotted] (\lineScale*\Pts, \ptsH) -- (\lineScale*\Pts, -\ptsH) node [at end, below] {$(i, \x)$};
  }
  \pgfmathsetmacro\Pts{Mod(7*\a,\modulus)}
  \draw (\lineScale*\Pts, \ptsH) -- (\lineScale*\Pts, -\ptsH) node [at end, below] {$(i+1, 0)$};

  \setcounter{cpt}{0}
  \foreach \x in {6, 6, 6, 6}{
    \pgfmathsetmacro\pt{\value{cpt}+(\x*0.5)}
    \node [anchor=south] at (\lineScale*\pt, 0) {\small $\theta_i$};
    \pgfmathsetmacro\ptStart{\value{cpt}}
    \addtocounter{cpt}{\x}
    \pgfmathsetmacro\ptEnd{\value{cpt}}
    \draw [->, in=120, out = 60, looseness=1, densely dashed] (\lineScale*\ptStart, 0) to (\lineScale*\ptEnd, 0);
  }
  
  \pgfmathsetmacro\ptStart{\value{cpt}}
  \foreach \x in {6, 6, 6}{
    \pgfmathsetmacro\pt{\value{cpt}+(\x*0.5)}
    \node [anchor=south] at (\lineScale*\pt, 0) {\small $\theta_i$};
    \addtocounter{cpt}{\x}
  }
  
  \pgfmathsetmacro\x{3}
  \pgfmathsetmacro\pt{\value{cpt}+(\x*0.5)}
  \node [anchor=south] at (\lineScale*\pt, 0) {\small $\theta_{i+1}$};
  \pgfmathsetmacro\ptEnd{\value{cpt}}

  \draw [->, in=120, out = 60, looseness=1] (\lineScale*\ptStart, 0) to (\lineScale*\ptEnd, 0);

\end{tikzpicture}
    }\hspace{2em}
    \resizebox{0.45\linewidth}{!}{
      \begin{tikzpicture}[>=latex]
  \pgfmathsetmacro\a{6} 
  \pgfmathtruncatemacro\modulus{45} 
  
  \pgfmathsetmacro\ptsH{0.25} 
  \pgfmathsetmacro\lineScale{0.3} 

  \draw (0, 0) -- (\lineScale*\modulus, 0);
  \draw (\lineScale*0, \ptsH) -- (\lineScale*0, -\ptsH) 
  ;
  \draw (\lineScale*\modulus, \ptsH) -- (\lineScale*\modulus, -\ptsH) 
  ;

  \fill (\lineScale*20, 0) circle (\lineScale*0.25) node [below] {$b$};
  
  \foreach \x in {1,...,6}{
    \pgfmathsetmacro\Pts{Mod(\x*\a,\modulus)}
    \draw [densely dotted] (\lineScale*\Pts, \ptsH) -- (\lineScale*\Pts, -\ptsH) node [at end, below] {$(i, \x)$};
  }
  \pgfmathsetmacro\Pts{Mod(7*\a,\modulus)}
  \draw (\lineScale*\Pts, \ptsH) -- (\lineScale*\Pts, -\ptsH) node [at end, below] {$(i+1, 0)$};

  \setcounter{cpt}{0}
  \foreach \x in {6, 6, 6, 6, 6, 6, 6}{
    \pgfmathsetmacro\pt{\value{cpt}+(\x*0.5)}
    \node [anchor=south] at (\lineScale*\pt, 0) {\small $\theta_i$};
    \addtocounter{cpt}{\x}
  }
  
  \pgfmathsetmacro\x{3}
  \pgfmathsetmacro\pt{\value{cpt}+(\x*0.5)}
  \node [anchor=south] at (\lineScale*\pt, 0) {\small $\theta_{i+1}$};
  \pgfmathsetmacro\ptEnd{\value{cpt}}

  \draw [->, densely dashed, in=150, out = 30, looseness=0.75] (0, 0) to (\lineScale*\ptEnd, 0);

\end{tikzpicture}
    }
  }
  
  \subfloat[$i$ is odd]{
    \centering
    \resizebox{0.45\linewidth}{!}{
      \begin{tikzpicture}[>=latex]
  \pgfmathsetmacro\a{6} 
  \pgfmathtruncatemacro\modulus{45} 
  
  \pgfmathsetmacro\ptsH{0.25} 
  \pgfmathsetmacro\lineScale{0.3} 

  \draw (0, 0) -- (\lineScale*\modulus, 0);
  \draw (\lineScale*0, \ptsH) -- (\lineScale*0, -\ptsH) 
  ;
  \draw (\lineScale*\modulus, \ptsH) -- (\lineScale*\modulus, -\ptsH) 
  ;

  \fill (\lineScale*25, 0) circle (\lineScale*0.25) node [below] {$b$};
  
  \foreach \x in {1,...,4}{
    \pgfmathsetmacro\Pts{Mod(\modulus - \x*\a,\modulus)}
    \draw (\lineScale*\Pts, \ptsH) -- (\lineScale*\Pts, -\ptsH) node [at end, below] {$(i, \x)$};
  }
  \foreach \x in {5,...,6}{
    \pgfmathsetmacro\Pts{Mod(\modulus - \x*\a,\modulus)}
    \draw [densely dotted] (\lineScale*\Pts, \ptsH) -- (\lineScale*\Pts, -\ptsH) node [at end, below] {$(i, \x)$};
  }
  \pgfmathsetmacro\Pts{Mod(\modulus - 7*\a,\modulus)}
  \draw (\lineScale*\Pts, \ptsH) -- (\lineScale*\Pts, -\ptsH) node [at end, below] {$(i+1, 0)$};

  \setcounter{cpt}{\modulus}
  \foreach \x in {6, 6, 6, 6}{
    \pgfmathsetmacro\pt{\value{cpt}-(\x*0.5)}
    \node [anchor=south] at (\lineScale*\pt, 0) {\small $\theta_i$};
    \pgfmathsetmacro\ptStart{\value{cpt}}
    \addtocounter{cpt}{-\x}
    \pgfmathsetmacro\ptEnd{\value{cpt}}
    \draw [->, out=120, in = 60, looseness=1, densely dashed] (\lineScale*\ptStart, 0) to (\lineScale*\ptEnd, 0);
  }
  
  \pgfmathsetmacro\ptStart{\value{cpt}}
  \foreach \x in {6, 6, 6}{
    \pgfmathsetmacro\pt{\value{cpt}-(\x*0.5)}
    \node [anchor=south] at (\lineScale*\pt, 0) {\small $\theta_i$};
    \addtocounter{cpt}{-\x}
  }
  
  \pgfmathsetmacro\x{3}
  \pgfmathsetmacro\pt{\value{cpt}-(\x*0.5)}
  \node [anchor=south] at (\lineScale*\pt, 0) {\small $\theta_{i+1}$};
  \pgfmathsetmacro\ptEnd{\value{cpt}}

  \draw [->, out=120, in = 60, looseness=1] (\lineScale*\ptStart, 0) to (\lineScale*\ptEnd, 0);
  
\end{tikzpicture}
    }\hspace{2em}
    \resizebox{0.45\linewidth}{!}{
      \begin{tikzpicture}[>=latex]
  \pgfmathsetmacro\a{6} 
  \pgfmathtruncatemacro\modulus{45} 
  
  \pgfmathsetmacro\ptsH{0.25} 
  \pgfmathsetmacro\lineScale{0.3} 

  \draw (0, 0) -- (\lineScale*\modulus, 0);
  \draw (\lineScale*0, \ptsH) -- (\lineScale*0, -\ptsH) 
  ;
  \draw (\lineScale*\modulus, \ptsH) -- (\lineScale*\modulus, -\ptsH) 
  ;

  \fill (\lineScale*25, 0) circle (\lineScale*0.25) node [below] {$b$};
  
  \foreach \x in {1,...,6}{
    \pgfmathsetmacro\Pts{Mod(\modulus - \x*\a,\modulus)}
    \draw [densely dotted] (\lineScale*\Pts, \ptsH) -- (\lineScale*\Pts, -\ptsH) node [at end, below] {$(i, \x)$};
  }
  \pgfmathsetmacro\Pts{Mod(\modulus - 7*\a,\modulus)}
  \draw (\lineScale*\Pts, \ptsH) -- (\lineScale*\Pts, -\ptsH) node [at end, below] {$(i+1, 0)$};

  \setcounter{cpt}{\modulus}
  \foreach \x in {6, 6, 6, 6, 6, 6, 6}{
    \pgfmathsetmacro\pt{\value{cpt}-(\x*0.5)}
    \node [anchor=south] at (\lineScale*\pt, 0) {\small $\theta_i$};
    \addtocounter{cpt}{-\x}
  }
  
  \pgfmathsetmacro\x{3}
  \pgfmathsetmacro\pt{\value{cpt}-(\x*0.5)}
  \node [anchor=south] at (\lineScale*\pt, 0) {\small $\theta_{i+1}$};
  \pgfmathsetmacro\ptEnd{\value{cpt}}

  \draw [->, densely dashed, out=150, in = 30, looseness=0.75] (\lineScale*\modulus, 0) to (\lineScale*\ptEnd, 0);

\end{tikzpicture}
    }
  }

  \caption{Behaviour of Lefèvre (left) and the new (right) HR-case searches when
    $b$ is in an interval of length $\theta_{i}$ (solid lines) and when $b$ is
    in an interval of length $\theta_{i-1, t}$ (dashed lines). Each point is
    labelled by the index $(i,t)$ of the two-length configuration it is added
    in.}
  \label{fig:four_cases_lefevre}
  \label{fig:four_cases_new}
\end{figure*}
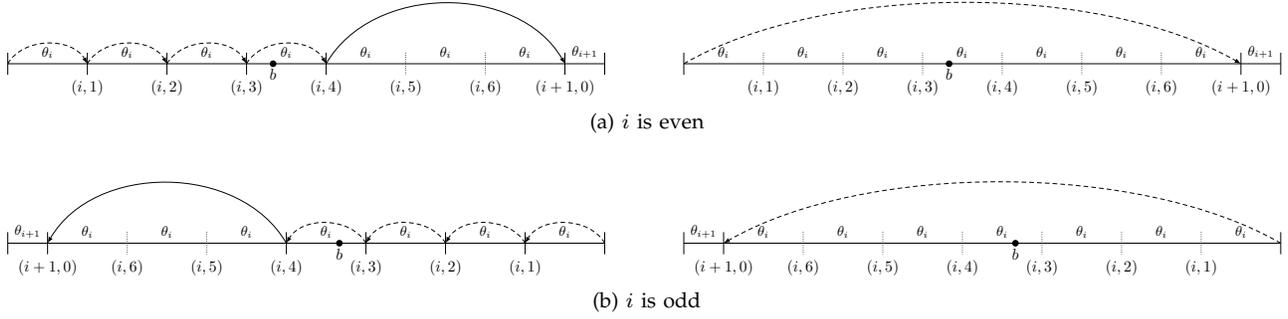

\alert{Hereafter we detail the relations between the two-length configurations
  and the execution paths of Algorithm \ref{algo_lefevre}. This algorithm starts
  with the configuration $(0, 1)$ and then considers the $(i,t)$
  configurations. It has to be noticed that the condition at line 4 is false
  only if we added one point directly at the left of $\{b\}$ during the previous
  loop iteration and true otherwise. Hence, it has to be interpreted as ``does
  $d$ need to be updated?''. This interpretation is allowed by the fact that the
  value of $d$ at line 4 corresponds to the previous configuration (the
  configuration $(0,0)$ at start) and that at least one point was already added
  (line 8 or 15). This condition enables thus to handle the next four cases
  (illustrated in Fig. \ref{fig:four_cases_lefevre})}.\alert{
\begin{itemize}
\item If $i$ is even:
  \begin{itemize}
  \item if $\{b\}$ is in an interval of length $\theta_i$, then \mbox{$d <
      \theta_i$} (this happens if the point previously added is just at the
    right of $\{b\}$). Hence no point is added in the interval containing
    $\{b\}$ so we go directly to the configuration $(i+1, 0)$ (lines 5-6) and
    $(i+1, 1)$ (line 8);
  \item if $\{b\}$ is in an interval of length $\theta_{i-1, t}$, then
    \mbox{$d>\theta_i$} (this case happens if the point previously added is just
    at the left of $\{b\}$). Hence $d$ is updated by subtracting $\theta_{i}$
    (line 10), $k = 0$ since $\theta_{i-1, t} > \theta_i$ (lines 12-13), and we
    go to configuration $(i, t+1)$ (line 15) as other points can be added to the
    left of $\{b\}$ in the next two-length configuration under Property
    \ref{prop:directed_reduction}.
\end{itemize}
\item If $i$ is odd:
  \begin{itemize}
  \item if $\{b\}$ is in an interval of length $\theta_i$, then \mbox{$d >
      \theta_{i-1, t}$} (this case happens if the point previously added is just
    at the left of $\{b\}$). Hence $d$ is updated (line 10) and we go to the
    configuration $(i+1, 0)$ (lines 12-13) and $(i+1, 1)$ (line 15);
  \item if $\{b\}$ is in an interval of length $\theta_{i-1,t}$, then
    \mbox{$d<\theta_{i-1,t}$} (this case happens if the point previously added
    is just at the right of $\{b\}$). Hence $k = 0$ since $\theta_{i-1, t} >
    \theta_i$ (lines 5-6) and we go to the configuration $(i, t+1)$ (line 8) as
    a point can be added to the left of $\{b\}$ in the next two-length
    configuration according to Property \ref{prop:directed_reduction}.
\end{itemize}
\end{itemize}
}
It has to be noticed that Lefèvre algorithm always reduces $d$ by using
subtractions at line 10 as points are added one by one at the left of
$\{b\}$. In practice Lefèvre adds specific instructions to compute partly these
reductions with divisions in order to avoid large quotients to be entirely
\alert{computed} with subtractions. We have omitted these instructions here for
clarity but they are present in our implementations of Lefèvre algorithm.

\label{sect_div_algo}
\alert{Furthermore the algorithm computes divisions (lines 5 and 12). In
  practice, we can make use of different division implementations. We can apply
  a subtractive division, a division instruction, or combine both in an hybrid
  approach as presented and analysed in \cite{Lefevre2005,Gouicem12}}.

\subsection{New regular HR-case search}
\label{sect_new_algo}
We here propose a new algorithm for the HR-case search where we use the same
filtering and division strategy as in Lefèvre algorithm, but we introduce a more
{\it regular} algorithm -- in the sense that it strongly reduces divergence on
GPU -- in order to compute a lower bound on \mbox{$\{b - a\cdot x
  \mod{1}~|~x<\#_pD_i\}$}. Hereafter, we will refer to this new algorithm as the
regular HR-case search.


\begin{algorithm}[t]
  \footnotesize
  \SetKwInOut{Input}{input}\SetKwInOut{Output}{output}
  \Input{$b-a\cdot x$, $\epsilon''$, $N$}
  {\bf initialisation: }
  \begin{tabular}{lll}
    $p \leftarrow \{a\}$; & $q \leftarrow 1$; & $d \leftarrow \{b\}$; \\
    $u \leftarrow 1$; & $v \leftarrow 0$; &
  \end{tabular}

  \lIf{ $d < \epsilon''$ }{\Return Failure\;}  
  \While {True}
  {
    \eIf{$p < q$}{
      $k = \lfloor q / p \rfloor$\;
      $q = q - k*p$; $u = u + k*v$\;
      $d = d \mod p$\;
    }
    {
      $k = \lfloor p / q \rfloor$\;
      $p = p - k*q$; $v = v + k*u$\;
      \If{$d \ge p$}{
        $d = (d - p) \mod q$\;
      }
    }
    \lIf{$u+v \ge N$}{\Return $ d > \epsilon''$\;}
  }
  \caption{New regular lower bound computation and test algorithm.}
  \label{algo_new}
\end{algorithm}

In this regular HR-case search described in Algorithm \ref{algo_new}, we only
consider configurations satisfying Equation \eqref{eq:config_div} in order to
use only the division-based Euclidean algorithm. The variables $p$ and $q$ store
respectively the lengths $\theta_i$ and $\theta_{i-1}$ for $i$ even, and the
lengths $\theta_{i-1}$ and $\theta_{i}$ for $i$ odd. The variables $u$ and $v$
store respectively $q_i$ and $q_{i-1}$ for $i$ even, and $q_{i-1}$ and $q_{i}$
for $i$ odd.

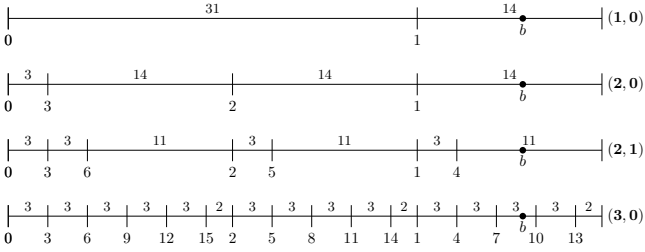
\begin{figure}[t]
  \centering
  \resizebox{\linewidth}{!}{
\begin{tikzpicture}
  \pgfmathsetmacro\a{31}
  \pgfmathsetmacro\b{39}
  \pgfmathtruncatemacro\modulus{45}
  
  \pgfmathsetmacro\ptsH{0.25} 
  \pgfmathsetmacro\whiteSpace{1.5} 
  \pgfmathsetmacro\lineScale{0.3} 
  \pgfmathsetmacro\blPos{\modulus + 0.05*\modulus} 
  \pgfmathsetmacro\nLabelPos{-1}
    
  \pgfmathsetmacro\N{0}
  
  \draw (0, -\whiteSpace*\N) -- (\lineScale*\modulus, -\whiteSpace*\N) node [at end, anchor = west] {$\mathbf{(1, 0)}$};
  \draw (\lineScale*0, -\whiteSpace*\N+\ptsH) -- (\lineScale*0, -\whiteSpace*\N-\ptsH) node [at end, below] {$0$};
  \draw (\lineScale*\modulus, -\whiteSpace*\N+\ptsH) -- (\lineScale*\modulus, -\whiteSpace*\N-\ptsH);
  
  \foreach \x in {0,...,1}{
    \pgfmathsetmacro\Pts{Mod(\x*\a,\modulus)}
    \draw (\lineScale*\Pts, -\whiteSpace*\N+\ptsH) -- (\lineScale*\Pts, -\whiteSpace*\N-\ptsH) node [at end, below] {$\x$} node [midway, above] {};
  }
  
  \setcounter{cpt}{0}
  \foreach \x in {31, 14}{
    \pgfmathsetmacro\pt{\value{cpt}+(\x*0.5)}
    \node [anchor=south] at (\lineScale*\pt, -\whiteSpace*\N) {\small $\x$};
    \addtocounter{cpt}{\x}
  }

  \fill (\lineScale*\b, 0) circle (\lineScale*0.25) node [below] {$b$};  
  
  \pgfmathsetmacro\N{1}
  
  \draw (0, -\whiteSpace*\N) -- (\lineScale*\modulus, -\whiteSpace*\N) node [at end, anchor = west] {$\mathbf{(2, 0)}$};
  \draw (\lineScale*0, -\whiteSpace*\N+\ptsH) -- (\lineScale*0, -\whiteSpace*\N-\ptsH) node [at end, below] {$0$};
  \draw (\lineScale*\modulus, -\whiteSpace*\N+\ptsH) -- (\lineScale*\modulus, -\whiteSpace*\N-\ptsH);
  
  \foreach \x in {0,...,3}{
    \pgfmathsetmacro\Pts{Mod(\x*\a,\modulus)}
    \draw (\lineScale*\Pts, -\whiteSpace*\N+\ptsH) -- (\lineScale*\Pts, -\whiteSpace*\N-\ptsH) node [at end, below] {$\x$} node [midway, above] {};
  }

  \setcounter{cpt}{0}
  \foreach \x in {3, 14, 14, 14}{
    \pgfmathsetmacro\pt{\value{cpt}+(\x*0.5)}
    \node [anchor=south] at (\lineScale*\pt, -\whiteSpace*\N) {\small $\x$};
    \addtocounter{cpt}{\x}
  }

  \fill (\lineScale*\b, -\whiteSpace*\N) circle (\lineScale*0.25) node [below] {$b$};

  \pgfmathsetmacro\N{2}
  
  \draw (0, -\whiteSpace*\N) -- (\lineScale*\modulus, -\whiteSpace*\N) node [at end, anchor = west] {$\mathbf{(2, 1)}$};
  \draw (\lineScale*0, -\whiteSpace*\N+\ptsH) -- (\lineScale*0, -\whiteSpace*\N-\ptsH) node [at end, below] {$0$};
  \draw (\lineScale*\modulus, -\whiteSpace*\N+\ptsH) -- (\lineScale*\modulus, -\whiteSpace*\N-\ptsH);
  
  \foreach \x in {0,...,6}{
    \pgfmathsetmacro\Pts{Mod(\x*\a,\modulus)}
    \draw (\lineScale*\Pts, -\whiteSpace*\N+\ptsH) -- (\lineScale*\Pts, -\whiteSpace*\N-\ptsH) node [at end, below] {$\x$} node [midway, above] {};
  }

  \setcounter{cpt}{0}
  \foreach \x in {3, 3, 11, 3, 11, 3, 11}{
    \pgfmathsetmacro\pt{\value{cpt}+(\x*0.5)}
    \node [anchor=south] at (\lineScale*\pt, -\whiteSpace*\N) {\small $\x$};
    \addtocounter{cpt}{\x}
  }

  \fill (\lineScale*\b, -\whiteSpace*\N) circle (\lineScale*0.25) node [below] {$b$};  

  \pgfmathsetmacro\N{3}
  
  \draw (0, -\whiteSpace*\N) -- (\lineScale*\modulus, -\whiteSpace*\N) node [at end, anchor = west] {$\mathbf{(3, 0)}$};
  \draw (\lineScale*0, -\whiteSpace*\N+\ptsH) -- (\lineScale*0, -\whiteSpace*\N-\ptsH) node [at end, below] {$0$};
  \draw (\lineScale*\modulus, -\whiteSpace*\N+\ptsH) -- (\lineScale*\modulus, -\whiteSpace*\N-\ptsH);
  
  \foreach \x in {0,...,15}{
    \pgfmathsetmacro\Pts{Mod(\x*\a,\modulus)}
    \draw (\lineScale*\Pts, -\whiteSpace*\N+\ptsH) -- (\lineScale*\Pts, -\whiteSpace*\N-\ptsH) node [at end, below] {$\x$} node [midway, above] {};
  }

  \setcounter{cpt}{0}
  \foreach \x in {3, 3, 3, 3, 3, 2, 3, 3, 3, 3, 2, 3, 3, 3, 3, 2}{
    \pgfmathsetmacro\pt{\value{cpt}+(\x*0.5)}
    \node [anchor=south] at (\lineScale*\pt, -\whiteSpace*\N) {\small $\x$};
    \addtocounter{cpt}{\x}
  }

  \fill (\lineScale*\b, -\whiteSpace*\N) circle (\lineScale*0.25) node [below] {$b$};

\end{tikzpicture}
  }
  \caption{Example where Algorithm \ref{algo_new} considers a
    configuration $(i+1, 1)$ for $i=1$.}
  \label{fig:tricky_case}
\end{figure}

Thus, instead of testing if $\{b\}$ went from a split interval to an unsplit one
like in Lefèvre HR-case search, we test here which length is reduced as in the
classical Euclidean algorithm, and then we reduce it and update $d$ accordingly
(as illustrated in Fig. \ref{fig:four_cases_new}). In practice, the quotients
are computed like in Lefèvre HR-case search with a subtractive division, a
division instruction or the hybrid approach. Now we detail the execution of the
algorithm. Let $(i,0)$ be a two-length configuration.
\begin{itemize}
\item If $i$ is even, then the test $p<q$ is true since $\theta_i <
  \theta_{i-1}$, we go to the configuration $(i+1, 0)$ (lines 5-6) and :
  \begin{itemize}
  \item if $\{b\}$ was in an interval of length $\theta_i$, no point was added
    in the interval containing $\{b\}$ and $d$ is not updated as $d<\theta_i$
    and \mbox{$d = d \mod{\theta_i}$} (line 7);
  \item if $\{b\}$ was in an interval of length $\theta_{i-1}$, points were
    potentially added to the left of $\{b\}$. Hence the distance $d$ is updated
    by reduction modulo $\theta_i$ (line 7) since intervals are split from the
    left under Property \ref{prop:directed_reduction}.
  \end{itemize}
\item If $i$ is odd, then the test $p<q$ (line 4) is false, we go to the
  configuration $(i+1, 0)$ (lines 9-10) and :
  \begin{itemize}
  \item if $\{b\}$ was in an interval of length $\theta_i$, no point was added
    in the interval containing $\{b\}$. However, we might subtract
    $\theta_{i+1}$ to $d$ if $d \ge \theta_{i+1}$ (line 12), which is similar to
    considering the configuration $(i+1, 1)$. Note that this would have been
    done in the next loop iteration at line 7 (see Fig. \ref{fig:tricky_case});
  \label{bad_case}
  \item if $\{b\}$ was in an interval of length $\theta_{i-1}$, points were
    potentially added to the left of $\{b\}$. According to Property
    \ref{prop:directed_reduction}, intervals are split from the right. Then the
    distance $d$ is updated if $d > \theta_{i+1}$ by reducing $d-\theta_{i+1}
    \mod \theta_i$ (lines 11-12).
  \end{itemize}
\end{itemize}

\subsection{Deployment on GPU}

\begin{figure}
  \centering
  \subfloat[Lefèvre HR-case search with specific instructions]{
    \centering
    \resizebox{0.8\linewidth}{!}{
      \includegraphics[width=\linewidth]{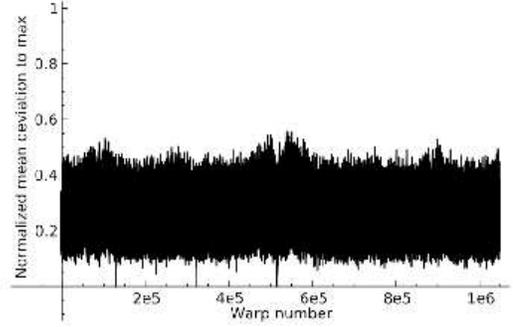}
    }
    \label{fig_nmdm_lefevre}
  }  

  \subfloat[Regular HR-case search]{
    \centering
    \resizebox{0.8\linewidth}{!}{
      \includegraphics[width=\linewidth]{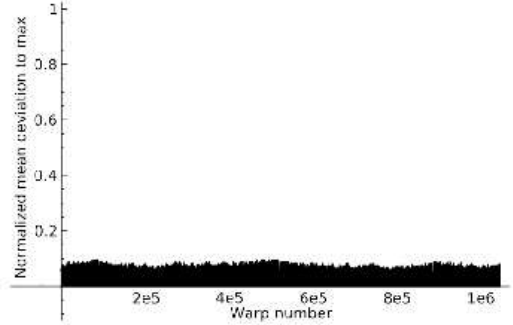}
    }
    \label{fig_nmdm_new}
  }
  \caption{Normalized mean deviation to the maximum of the number of main loop
    iterations per warp among the $2^{20}$ warps required for the \emph{exp}
    function in the domain $[1; 1+2^{-13}]$.}
  \label{fig_nmdm_comparison}
\end{figure}

The exhaustive search algorithm perfectly takes advantage of the GPU massive
parallelism and of its (partial) SIMD execution. Hence, we will focus on the
deployment of the lower bound computation.  In this section we present the GPU
deployment of Lefèvre HR-case search as detailed in \cite{Gouicem12}, and the
GPU deployment of the new regular HR-case search. We particularly study the
divergence in both HR-case searches at three levels: the filtering strategy, the
main loop and the main conditional statement. For these deployments, we first
changed the data layout to a ``structure of arrays'' in order to have coalesced
memory accesses \cite[Sect.~6.2.1]{nvidia_best_practice}. We also avoided as
much as possible consecutive dependent instructions in order to increase the
instruction-level parallelism within each thread.

Throughout this section, we will consider the example domain $[1,
1+2^{-13}[$ in the binade $[1,2[$ for the exponential function in double
precision, as this binade is considered in \cite{Lefevre2005} as the general
case.

\subsubsection{Filtering strategy divergence}
\label{sect_filtering}

As a consequence of the filtering strategy, we will have few threads executing
phase 2, and fewer executing phase 3. Table \ref{table_step_num_interv} shows
the number of sub-domains involved in each phase for a domain $D_i$ containing
$2^{40}$ floating-point numbers. As we can see, very few sub-domains
are concerned by the exhaustive search step. Hence, executing one kernel
computing the three phases leads to an important divergence as we have fewer and
fewer active threads within each warp from one phase to the next
\cite{Gouicem12}.

To tackle this problem, we propose to use three kernels, one for each
phase. This allows us to re-build the grid of threads between each phase, and to
run the exact number of threads required by each phase.
However, this implies two additional costs. 

First, we have to write failing sub-domains\footnote{Sub-domains for which the
  computed lower bound is less than $\epsilon''$ in algorithms
  \ref{algo_lefevre} and \ref{algo_new}.} of phases 1 and 2 in consecutive
memory locations as we prepare coalesced reads for the next phase.  In
\cite{Gouicem12}, this was done with atomic operations on the GPU global memory
since we had few failing sub-domains. For some specific binades, the number of
failing sub-domains can be much more important and the numerous atomic
operations can then lower the performance. Hence, we use atomic operations on
the GPU global memory or compact operations based on parallel prefix sums
provided by CUDPP \cite{Sengupta2008}, depending on the expected number of
failing sub-domains.

Second, between two phases, we have to transfer back to CPU the number of
failing sub-domains to compute on CPU the optimal grid size for the next
phase. 


\begin{table}
  \centering
    \begin{tabular}{ccc}
      \cline{2-3}
      \multirow{2}{*}{} & \multicolumn{2}{c}{Number of arguments} \\
      & Lefèvre & Regular \\
      \hline
      Phase 1 & $2^{40} \approx 1.1\cdot 10^{12}$ & $2^{40} \approx 1.1\cdot 10^{12}$ \\
      Phase 2 & $\approx 3.6\cdot 10^{9}$ & $ \approx 1.8\cdot 10^{10}$ \\
      Phase 3 & $\approx 8.9\cdot 10^6$ & $\approx 5.9\cdot 10^7$ \\
      \hline
      HR-cases & $243$ & $243$\\
      \hline
    \end{tabular}
    \caption{Details of argument filtering during HR-case search in
      $[1, 1+2^{-13}]$.}
  \label{table_step_num_interv}
\end{table}


It can be noticed in Table \ref{table_step_num_interv} that Lefèvre HR-case
lower bound computation filters a little more than the new algorithm. Lefèvre
HR-case lower bound computation uses indeed subtraction-based Euclidean
algorithm when splitting the interval containing $\{b\}$. This results in a
number of considered arguments less than $2N$. On the contrary, we always use
the division-based Euclidean algorithm in the regular HR-case search. If we
consider $i$ such that $q_i + q_{i-1} < N < q_{i+1} + q_i$, then $q_{i+1} + q_i
= q_{i-1} + k_{i+1}q_i + q_i < (k_{i+1}+1)N$ -- by considering $q_i <
N$. However, the geometric mean of the quotients $k_i$ of the continued fraction
of almost all real numbers equals Khinchin's constant ($\approx 2.69$)
\cite{Khinchin92}. Hence, using the regular HR-case search we consider on
average less than $3.69\cdot N$ arguments.

\subsubsection{Loop divergence}

\begin{table}
  \centering

  \begin{tabular}{lcccc}
    \hline
    \multirow{3}{*}{HR-case search} & min & max  & mean & mean\\
    & iteration & iteration & iteration & NMDM\\
    & number & number & number & \\
    \hline
    Lefèvre & $5$ & $328$ & $24$ & $25.6\%$\\
    with specific & \multirow{2}{*}{$5$} & \multirow{2}{*}{$31$} & \multirow{2}{*}{$16$} & \multirow{2}{*}{$25.7\%$}\\ 
    instructions &&&&\\
    Regular & $8$ & $19$ & $12$ & $0.1\%$\\
    \hline
  \end{tabular}
  
  \caption{Comparison of the main loop behavior among the $2^{20}$ warps required for the different HR-case searches on \emph{exp} function in the domain $[1, 1+2^{-13}]$.}
  \label{table_nmdm_comparison}
\end{table}

The second source of divergence is the main unconditional loop (see line 3 in
Algorithms \ref{algo_lefevre} and \ref{algo_new}). Fig.
\ref{fig_nmdm_comparison} shows the NMDM of the number of loop iterations by
warp for the different HR-case searches when testing a domain $D_i$
containing $2^{40}$ double precision floating-point arguments. Table
\ref{table_nmdm_comparison} summarizes statistical informations on the NMDM and
the number of iterations for both Lefèvre and the regular HR-case searches.

For Lefèvre HR-case search, this main unconditional loop is an important source
of divergence with a mean NMDM of $25.6\%$, that is to say, a thread remains
idle on average $25.6\%$ of the number of loop iterations executed by its
warp. To our knowledge there is no \emph{a priori} information on the number of
loop iterations that would enable us to statically reorder the sub-domains in
order to decrease this divergence. We also tried to use software solutions to
reduce the impact of the loop divergence \cite{Gouicem12} to no avail because
the computation is very fine-grained.

This divergence in Lefèvre HR-case search is mainly due to the fact that the
quotients are entirely or partially computed at each iteration depending on the
position of $b$ even with the specific instructions (see
Sect. \ref{sect_lefevre_algo}). Thanks to these specific instructions the
pathological cases are avoided (see Table \ref{table_nmdm_comparison}) but the
mean NMDM is still around $25.6\%$.

In the new regular HR-case search, the key point is that a quotient of the
continued fraction expansion of $a$ is entirely computed at each loop iteration,
which is not the case in Lefèvre HR-case search. Hence, the number of loop
iterations only depends on the number of quotients of the continued fraction
expansion of $a$ computed to reach $\#_pD_i$ points on the segment. As the
number of quotients to compute is very close from one sub-domain to the next, we
reduce the mean NMDM by warp to $0.1\%$.

\subsubsection{Branch divergence}

The third source of divergence is on the main conditional statement (see line 4
in Algorithms \ref{algo_lefevre} and \ref{algo_new}). We aim at reducing the
number of instructions controlled by the branch condition, and if reduced
enough, benefit from the GPU branch predication
\cite[Sect.~9.2]{nvidia_best_practice}. This branch predication enables indeed,
for short sections of divergent code, to fill at best the pipelines by
scheduling both {\it then} and {\it else} branches for all threads: thank to a
per-thread predicate, only the relevant results are actually computed and
finally written.


\begin{algorithm}[t]
  \footnotesize
  \SetKwInOut{Input}{input}
  \SetKwInOut{Output}{output}
  \Input{$b - a\cdot x$, $\epsilon''$, $N$}
  {\bf initialisation: }
  \begin{tabular}{lll}
    $p \leftarrow \{a\}$; & $q \leftarrow 1 - \{a\}$; & $d \leftarrow \{b\}$; \\
    $u \leftarrow 1$; & $v \leftarrow 1$; & $are\_swapped \leftarrow False$;
  \end{tabular}

  \lIf{ $d < \epsilon''$ }{\Return Failure\;}
  \If{$(d \ge p)$}
  {
    SWAP$(p,q)$; SWAP$(u,v)$\;
    $are\_swapped \leftarrow True$\;
  }
  
  \While {True}
  {
    \If{$are\_swapped$}{
      $d \leftarrow d - p$\;
      \lIf{$d < \epsilon''$}{\Return Failure\;}
    }
    
    $k = \lfloor q/p \rfloor$\;
    $q \leftarrow q - k*p$; $u \leftarrow u+ k*v$\;
    \lIf{$u+v \ge N$}{\Return Success\;}
    $p \leftarrow p - q$; $v \leftarrow v + u$\;
    
    \If{$are\_swapped$ xor $(d \ge p)$}
    {
      SWAP$(p,q)$; SWAP$(u,v)$\;
      $are\_swapped \leftarrow not(are\_swapped)$\;
    }
  }
  \caption{Lefèvre's lower bound computation and test algorithm with swap.}
  \label{algo_lefevre_swap}
\end{algorithm}

As observed in \cite{Gouicem12}, both branches of Lefèvre HR-case search contain
the same instructions, except that the variables $p$ (respectively $u$) and $q$
(resp. $v$) are interchanged, and that $p$ is subtracted to $d$ in the {\it
  else} branch.  We therefore swap the two values $p$ and $q$ (resp. $u$ and
$v$) to remove the common instructions from the conditional scope as described
in Algorithm \ref{algo_lefevre_swap}. The swap implies a small extra cost but we
thus reduce the number of divergent instructions.


\begin{algorithm}[t]
  \footnotesize
  \SetKwInOut{Input}{input}\SetKwInOut{Output}{output}
  \Input{$b - ax$, $\epsilon''$, $N$}
  {\bf initialisation: }
  \begin{tabular}{lll}
    $p \leftarrow \{a\}$; & $q \leftarrow 1$; & $d \leftarrow \{b\}$; \\
    $u \leftarrow 1$; & $v \leftarrow 0$; &
  \end{tabular}
  
  \While {True}
  {
    $k = \lfloor q / p \rfloor$\;
    $q = q - k*p$;
    $u = u + k*v$\;
    $d = d \mod p$\;
    \lIf{$u+v \ge N$}{\Return $ d > \epsilon''$\;}

    $k = \lfloor p / q \rfloor$\;
    $p = p - k*q$;
    $v = v + k*u$\;
    \If{$d \ge p$}{
      $d = d - p \mod q$\;
    }
    \lIf{$u+v \ge N$}{\Return $ d > \epsilon''$\;}
  }
\caption{New regular lower bound computation and test algorithm
  unrolled.}
\label{algo_new_unrolled}
\end{algorithm}

As far as the new regular HR-case search is concerned, there is in Algorithm
\ref{algo_new} as much branch divergence within the unconditional
loop as in Algorithm \ref{algo_lefevre}. However the main conditional statements
of the two algorithms are rather different. In Lefèvre HR-case search, this test
depends on the position of the point $b$ at each iteration. In the regular
HR-case search, it depends on the length to reduce. Unlike the test on the
position of $b$, the test on the length to reduce is deterministic as the
regular HR-case search computes a quotient of the continued fraction expansion
of $a$ at each loop iteration.  Hence the evaluation of the condition switches
at each loop iteration and it first evaluates to {\it True} as $p$ is
initialized to $\{a\}$ and $q$ to $1$. Therefore, by unrolling two loop
iterations (Algorithm \ref{algo_new_unrolled}), we can avoid this
test and strongly reduce the branch divergence.


\section{Polynomial approximation generation on GPU}
\label{sect_pol_approx}

In this section, we detail how we have deployed on GPU the generation of the
polynomial approximations $P_i$ required for the HR-case search algorithms
described in Sect. \ref{sect_hr_search}. \alert{We recall that the change of
  variable \mbox{$x = 2^{p-e(X_i)}(X-X_i)$} enables to test the floating-point
  arguments $X\in D_i$ of $f(X)$ by testing the integer arguments \mbox{$x\in
  \llbracket 0, \#_pD_i - 1 \rrbracket$} of $P_i(x)$}.

\alert{ Computing as many approximations as $D_i$ domains can be prohibitive
  using Taylor approximations. The principle here is therefore to consider the
  union of $\tau$ domains $D_t, \dots, D_{t+\tau-1}$ -- denoted $\mathcal{D}_t$
  -- and to approximate the function $f$ by a polynomial $R_t$ of degree
  $\delta$ -- with a Taylor approximation for example -- such that $| R_t(x) -
  f(X) | < \epsilon_{approx}'2^{e(f(X)) - p}$ for all $X\in \mathcal{D}_t$ with
  $x = 2^{p-e(X_t)}(X-X_t)$}.

\alert{If $\tau$ is chosen such that $e(x) = e(y)$ for all
  $x,y\in\mathcal{D}_t$, then $P_{t+i}(x)$ is defined as $R_t(x+iN)$ for $0 \le
  i < \tau$ with $N=\#_pD_t$ (as $\#_pD_i = \#_pD_j~,~\forall i, j \in
  \mbox{$\llbracket t, t+\tau-1 \rrbracket$}$). The shifts of the form
  $R_t(x+iN)$ are called \emph{Taylor shifts} \cite{vonzurGathen97}. If we
  denote $\epsilon_{shift}$ the error propagated by the shift such that
  $|P_{t+i}(x) - R_t(x + iN)| < \epsilon_{shift}$, then we set
  $\epsilon_{approx}$ to $\epsilon_{approx} = \epsilon_{approx}' +
  \epsilon_{shift}$ (see Sect. \ref{sect:tmd})}.

In the following, we want to compute these polynomials $P_{t+i}$\;. We first
present a method named the \emph{hierarchical} method \alert{originally designed
  by Lefèvre \cite{Lefevre_thesis}} to change one Taylor shift by $N$ into
several Taylor shifts by $1$. Then, we present two existing Taylor shift
algorithms:
\begin{itemize}
\item the \emph{tabulated difference shift} which, starting with \mbox{$P_t(x) =
  R_t(x)$}, sequentially iterates a shift of the polynomial $P_{t+i}$ to obtain
  $P_{t+i+1}$ with only multi-precision additions \alert{\cite{dinechin2011}};
\item and the \emph{straightforward shift} which computes the $P_{t+i}$'s from
  $R_t$ in parallel but requires multi-precision multiplications and additions
  \alert{\cite{vonzurGathen97}}.
\end{itemize}
Finally we propose an hybrid CPU-GPU Taylor shift algorithm which efficiently
combines these two shifts with the hierarchical method, and which requires only
fixed size multi-precision addition on GPU.
More details on these algorithms and their error propagation can be found in
\cite{dinechin2011}.

\subsection{Hierarchical method}

We first describe the hierarchical method originally described in
\cite{Lefevre_thesis} which transforms one shift by $N$ of a polynomial
\alert{$R_t(x)$} of degree $\delta$ into $\delta+1$ shifts by $1$. This is of
interest as shifting by $1$ can be done with only additions (see
Sect. \ref{sec:tayl-shift-algor}).  This method requires the input polynomial to
be interpolated in the binomial basis
$\binom{x}{j}=\frac{\prod_{l=0}^{j-1}{(x-l)}}{j!} $. Therefore, we define the
forward difference operator and its application to interpolate a polynomial in
the binomial basis.

\begin{definition}
  The forward difference operator, denoted $\Delta_h$ is defined as
  $\Delta_h[P](x) = P(x + h) - P(x)$. \alert{We write $\Delta_h^j$ the
    composition $j$ times} of $\Delta_h$ and $\Delta = \Delta_1$.
\end{definition}

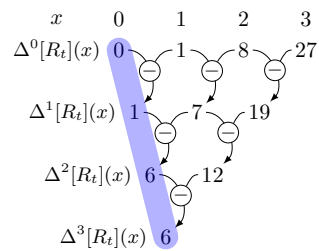
\begin{figure}
  \centering
  \resizebox{0.5\linewidth}{!}{

\begin{tikzpicture}[>=latex]
    \pgfmathsetmacro\LineShift{0.25}
    \pgfmathsetmacro\PlusNodeShiftx{0.5}
    \pgfmathsetmacro\PlusNodeShifty{0.4}

    \pgfmathtruncatemacro\NbLine{3}
    \pgfmathtruncatemacro\NbCol{3}
    \pgfmathtruncatemacro\NbLineMinusOne{\NbLine - 1}
    \pgfmathtruncatemacro\NbColMinusOne{\NbCol - 1}

    \def\Values{{{0,1,8,27}, {1,7,19}, {6,12}, {6}}}

    \node at (-1, 0.5) {$x$};
    \foreach \x in {0,...,\NbCol}
    {
      \node[inner sep=0.5mm] at (\x, 0.5) {$\x$};
    }

    \foreach \y in {0,...,\NbCol}
    {
      \node[inner sep=0.5mm] at (-1 + \y*\LineShift, -\y) {\small $\Delta^\y[R_t](x)$};
    }

    \foreach \y in {0,...,\NbLine}
    {
      \pgfmathtruncatemacro\NbColLine{\NbCol-\y}
      \foreach \x in {0,...,\NbColLine}
      {
        \node[inner sep=0.5mm] at (\x + \y*\LineShift, -\y) (d_\x_\y) {\pgfmathparse{\Values[\y][\x]}$\pgfmathresult$};
      }
    }

    \foreach \y in {1,...,\NbLine}
    {
      \pgfmathtruncatemacro\NbColLine{\NbCol-\y}
      \foreach \x in {0,...,\NbColLine}
      {
        \pgfmathtruncatemacro\X{\x+1}
        \pgfmathtruncatemacro\Y{\y-1}
        \draw[->] (d_\x_\Y) to[out =0,in=45, looseness=1] 
           node[draw, circle, inner sep=0mm, pos=0.5, fill=white] (m_\x_\y) {\small $-$} (d_\x_\y);
        \draw (d_\X_\Y) to[out=180, in=60, looseness=1] (m_\x_\y);
      }
    }

    \pgfmathtruncatemacro\yStart{-0.1}
    \pgfmathtruncatemacro\yEnd{\NbLine+0.1}
    \draw[color=blue, opacity=.3,line width=4mm, line cap=round] (0 - \yStart*\LineShift, -\yStart) to (0 + \yEnd*\LineShift, -\yEnd);
          
  \end{tikzpicture}
  }
  \caption{Newton interpolation of polynomial $x^3$. The coefficients of
    the interpolated polynomial are highlighted.}
  \label{fig_newton_interpolation}
\end{figure}

\begin{figure}
  \centering
  \resizebox{0.7\linewidth}{!}{

  \begin{tikzpicture}
    \pgfmathsetmacro\LineShift{0.25}
    \pgfmathsetmacro\PlusNodeShiftx{0.5}
    \pgfmathsetmacro\PlusNodeShifty{0.4}

    \pgfmathtruncatemacro\NbLine{3}
    \pgfmathtruncatemacro\NbCol{5}
    \pgfmathtruncatemacro\NbLineMinusOne{\NbLine - 1}
    \pgfmathtruncatemacro\NbColMinusOne{\NbCol - 1}

    \def\Values{{{0,1,8,27, 64, 125, 216}, {1,7,19, 37, 61, 91, 0}, {6,12,18,24,30,36,0}, {6,6,6,6,6,6}}}

    \foreach \y in {0,...,\NbLine}
    {
      \foreach \x in {0,...,\NbCol}
      {
        \node[inner sep=0.5mm] at (\x + \y*\LineShift, -\y) (d_\x_\y) {\pgfmathparse{\Values[\y][\x]}$\pgfmathresult$};
      }
    }

    \foreach \y in {0,...,\NbLine}
    {
      \node[inner sep=0.5mm] at (-1 + \y*\LineShift, -\y) {\small $\Delta^\y[r_{t,j}](i)$};
    }
    
    \foreach \y in {0,...,\NbLineMinusOne}
    \foreach \x in {0,...,\NbColMinusOne}
    {
      \pgfmathtruncatemacro\X{\x+1}
      \pgfmathtruncatemacro\Y{\y+1}
      \draw[->] (d_\x_\y) to[out =-45,in=-135, looseness=1] 
      node[draw, circle, inner sep=0mm, pos=0.5, fill=white] (p_\x_\y) {\small $+$} (d_\X_\y);
      \draw (d_\x_\Y) to[out=90, in = -135, looseness=1] (p_\x_\y);      
    }

    \foreach \x in {0,...,\NbCol}
    {
      \pgfmathtruncatemacro\yStart{-0.1}
      \pgfmathtruncatemacro\yEnd{\NbLine+0.1}
      \draw[color=blue, opacity=.\x,line width=4mm, line cap=round] (\x - \yStart*\LineShift, -\yStart) to (\x + \yEnd*\LineShift, -\yEnd);
    }
          
  \end{tikzpicture}
  }
  \caption{Tabulated difference shift for evaluating the polynomial
    $r_{t,j}(i)=i^3$. The shifted polynomials are highlighted.}
  \label{fig_tabulated_difference}
\end{figure}
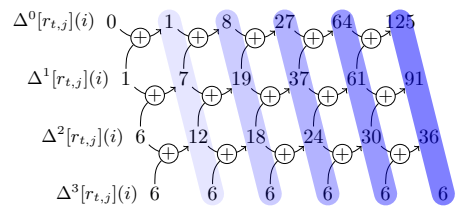

Using this forward difference operator, one can efficiently interpolate
\alert{the polynomial $R_t$ of degree $\delta$ in the binomial basis
  \cite{dinechin2011}, given the values $\{ R_t(x), x\in\llbracket 0,\delta
  \rrbracket \}$ as $R_t(x) = \sum_{j=0}^{\delta}\Delta^j[R_t](0)\cdot
  \binom{x}{j}.$} An example is shown in Fig. \ref{fig_newton_interpolation}.
This interpolation is computed using the definition of $\Delta$ and with initial
values \alert{$\Delta^0[R_t](x) = R_t(x)$}.  This algorithm is similar to the
Newton interpolation with the forward difference operator used instead of the
forward divided difference operator.

Now, we describe the hierarchical method\alert{\cite{Lefevre_thesis}}. Given a
polynomial \alert{$R_{t}(x)$}, we want to build a scheme to shift this
polynomial in consecutive arguments following an arithmetic progression with
common difference $N$. Let us consider the univariate polynomial \alert{$R_t$ as
  a bivariate polynomial \mbox{$R_t(x + iN)$} in the variables $x$ and $i$. By
  interpolation in the binomial basis with respect to the variable $x$, we
  obtain a polynomial in the variable $x$, with polynomial coefficients
  $r_{t,j}(i) = \Delta^j[R_t](iN)$ in the variable $i$, defined as follow
  \begin{equation*}
    R_{t}(x + iN) = \sum_{j=0}^\delta{r_{t,j}(i)\binom{x}{j}}.  
  \end{equation*}}

\alert{Using the hierarchical method, we thus obtain the polynomials
  $r_{t,j}(i)$. From these, one can compute the shifts $P_{t+i}(x)$ of the
  polynomial $R_{t}(x)$ by $iN$ by computing the evaluations $r_{t,j}(i)$ with
  $0 \le j \le \delta$}. If we consider the polynomials $r_{t,j}$ in the
binomial basis, these evaluations $r_{t,j}(i)$ can be obtained with the
consecutive Taylor shifts of $r_{t,j}$ by one -- by taking the coefficients of
degree $0$ of these shifts, which are the $\Delta^0[r_{t,j}](i)$.

\subsection{Taylor shift algorithms}
\label{sec:tayl-shift-algor}
\alert{Taylor shifts by one can be performed efficiently with the {\it tabulated
    difference} shift \cite{dinechin2011, Lefevre1998}}. According to the
forward difference operator definition, \mbox{$\Delta^l[r_{t,j}](i)
  =\Delta^{l-1}[r_{t,j}](i+1) - \Delta^{l-1}[r_{t,j}](i),$} that is to say
\mbox{$\Delta^{l-1}[r_{t,j}](i+1) = \Delta^{l-1}[r_{t,j}](i) +
  \Delta^l[r_{t,j}](i).$} Furthermore, if $deg(r_{t,j}) = \gamma$ then
\alert{$\Delta^{\gamma}[r_{t,j}](i)$} is constant for any integer \alert{$i \ge
  0$}, as it is the $\gamma^{th}$ discrete derivative of \alert{$r_{t,j}$} times
$\gamma!$\,. An illustration of this algorithm can be found in Fig.
\ref{fig_tabulated_difference}. Hence, the only needed operations to obtain the
consecutive evaluations of the polynomials \alert{$r_{t,j}$} are multi-precision
additions of the coefficients.  

Obtaining the consecutive evaluations of \alert{$r_{t,j}$} can also be performed
with the {\it straightforward shift}. This algorithm multiplies \alert{the
  vector of the $r_{t,j}$ polynomial coefficients} by a matrix constructed using
Newton's binomial theorem. If we consider the polynomials $r_{t,j}$
  expressed in the binomial basis, this multiplication exactly corresponds to
  applying $i$ times the tabulated difference algorithm on the polynomial
  $r_{t,j}$. This matrix is upper triangular and Toeplitz, which can be used to
  speed up the matrix-vector multiplication for high degree, and is constructed
  as \[
  \small
\begin{pmatrix}
  \binom{i}{0}  & \binom{i}{1} & \cdots & \binom{i}{\gamma} \\
  0             & \binom{i}{0} & \cdots & \binom{i}{\gamma-1} \\
  \vdots             & \ddots             & \ddots & \vdots \\
  0             & \cdots        & 0      & \binom{i}{0}
\end{pmatrix}.
\]
Therefore, to construct this matrix, only the first $\binom{i}{l}$ with $0
\le l <\gamma+1$ are needed to compute its coefficients.

\subsection{Hybrid CPU-GPU deployment}
\alert{Now, we propose a hybrid CPU-GPU deployment of the polynomial
  approximation generation step}. The \alert{polynomials $R_t$ are Taylor
  polynomials} of degree $\delta=2$ approximating the targeted function over
$\tau=2^{25}$ domains $D_i$ like in \cite{Lefevre1998}. We interpolate
\alert{them} in the binomial basis using the hierarchical method with $N=2^{15}$
as we want to use the Boolean tests described in Sect.  \ref{sect_hr_search} on
intervals containing $2^{15}$ arguments (\alert{this choice of parameters is
  motivated by the error analysis of the Boolean test
  \cite{Lefevre_thesis}}). \alert{More formally, we have $$P_{t+i}(x)=
  R_t(i2^{15}+x) = \sum_{j=0}^{2}{r_{t,j}(i)\binom{x}{j}}.$$} As this
interpolation in the binomial basis is done once, it is precomputed on CPU.


Hence, to obtain all \alert{the polynomials $P_{t+i}$ for \mbox{$0 \le i <
    \tau$}}, we have to deploy on GPU the computation of the consecutive
evaluations of \alert{$r_{t,j}(i)$ for $0 \le i < \tau$ and $0 \le j \le 2$}.

On one hand, the tabulated difference shift is efficient as it requires only
multi-precision additions. This method is thus used in the reference CPU
implementation \cite{Lefevre1998}. However this is an intrinsically sequential
\alert{algorithm}, which prohibits its direct deployment on GPU.
On the other hand, the straightforward shift is embarrassingly parallel, but
requires multi-precision multiplications and divisions to compute the binomials
and multi-precision multiplications and additions to compute the matrix-vector
products.

In order to benefit from the efficiency of the tabulated difference shift on
GPU, we therefore use an hybrid strategy that relies on both the CPU and the
GPU: we compute \alert{the shifts $r_{t,j, u}(i) = r_{t,j}(u\nu + i)$ to form
  $\mu$ packets of size $\nu$ such that $\mu\nu = \tau$. We vary $u$ from $0$ to
  $\mu - 1$ and construct the polynomials $r_{t, j, u}$ sequentially on CPU with
  the straightforward shift}\footnote{This computation on CPU could thus be
  parallelized but the corresponding computation times are minority in
  practice.}. All the multi-precision operations on CPU are computed efficiently
using the GMP library \cite{GMP}.

Then the \alert{ $\mu$ polynomials $r_{t,j, u}$} are transferred to GPU. We run
a CUDA kernel of \alert{$\mu$} threads wherein each thread of ID $u$ processes
\alert{the polynomial $r_{t,j, u}$ and computes the evaluations $r_{t,j, u}(i)$
  with $0\le i<\nu$ using the tabulated difference shift}.

Furthermore, as there are $\delta +1$ independent $r_{t,j}$ polynomials ($\delta
= 2$ in practice), we can run one kernel per \alert{$r_{t,j}$ polynomial} and
overlap the GPU tabulated difference shift for the polynomial \alert{$r_{t,j}$}
with the CPU straightforward shift of the polynomial \alert{$r_{t, j+1}$}. The
only algorithm deployed on GPU is therefore the tabulated difference shift which
is sequential within each GPU thread, but performed concurrently by multiple
threads on multiple polynomials \alert{$r_{t, j, u}$}.

As the coefficients of the considered polynomials are large, we need
multi-precision addition on GPU. Here only fixed size multi-precision additions
are required as bounds on the required precision, depending on the targeted
function and exponent of the targeted domain, can be computed before compile
time \cite{Lefevre_thesis, dinechin2011}. Multi-precision libraries on GPU
\cite{CUMP, GARPREC} have been very recently developed. However, we preferred to
have our own implementation of this operation for two main reasons: to use PTX
(NVIDIA assembly language) \cite{nvidia_ptx} and the \emph{addc} instruction in
order to have an efficient carry propagation; and to benefit from the fixed size
of the multi-precision words at compile time in order to unroll inner loops. As
the \emph{addc} instruction operates only on 32-bit words, multi-precision words
are arrays of 32-bit chunks. The multi-precision addition function is
implemented as a C++ template with the size of the multi-precision words given
as a parameter, which enables an automatic generation of addition functions for
each size of fixed multi-precision word required by each binade. As a
consequence, the inner loop on the number of chunks can easily be unrolled as
the number of loop iterations is known at compile time. Furthermore, in order
to have coalesced memory accesses, the word chunks are interleaved in global
memory and loaded chunk by chunk in registers. 

Finally, it can be noticed that this algorithm is completely regular: there is
therefore no divergence issue among the GPU threads here.


\section{Performance results}
\label{sect_perf}

In this section we present the performance analysis of our different
deployments.  All results are obtained on a server composed of one Intel Xeon
X5650 hex-core processor running at 2.67 GHz, one NVIDIA Fermi C2070 GPU and 48
GB of DDR3 memory.

We compare three implementations. The first one is the sequential implementation
(named \emph{Seq.}) which is Lefèvre reference code provided by V. Lefèvre. The
second one is the parallel implementation on CPU (referred to as \emph{MPI})
which is the sequential implementation with an MPI layer (OpenMPI version 1.4.3)
to distribute equally the $2^{13}$ intervals composing a binade among the
available CPU cores. We use a cyclic decomposition which offers a better load
balancing than a block decomposition and run 12 MPI processes to take advantage
of the two-way SMT (Simultaneous Multithreading or Hyper-Threading for Intel) of
each core. The third implementation (named \emph{CPU-GPU}) relies on the GPU and
CPU-GPU deployments presented in this paper. The implementations have been
compiled with gcc-4.4.5 for CPU code and nvcc (CUDA 4.1) for GPU code.

All the following timings are obtained for searching $(53,2^{-32})$ HR-cases of
\emph{exp} function, that is to say double precision floating-point arguments
for which $32$ extra bits of precision during evaluation do not suffice to
guarantee correct rounding. The measures include all computations and data
transfers between the GPU and the CPU. \alert{All the tested implementations
  return the same HR-cases in the considered binades}.

\subsection{HR-case search}

We first searched for the optimal block sizes on GPU and tried to increase the
number of intervals computed per thread in every GPU kernel, in order to
optimize occupancy and computation granularity. However increasing the number of
intervals per thread do not improve performances since the occupancy of each
kernel is already high enough.  

\begin{table}
  \centering

    \begin{tabular}{cccc|cc}
      \cline{2-6}
      \multicolumn{1}{r}{} & {Seq.} & {MPI} & {CPU-GPU} & $\frac{\text{Seq.}}{\text{MPI}}$ & $\frac{\text{Seq.}}{\text{CPU-GPU}}$ \\
      \hline
      {Pol.} & \multirow{2}{*}{$43300.81$} & \multirow{2}{*}{$5251.53$} & \multirow{2}{*}{$788.84$} & \multirow{2}{*}{$8.25$} & \multirow{2}{*}{$54.89$}\\
      approx. &&&&&\\
      {Lefèvre}& $36816.10$ & $5292.67$ & $2446.27 $ & $6.96$ & $15.05$\\
      {Regular} & $34039.94$ & $4716.97$ & $711.92$ & $7.22$ & $47.81$\\
      \hline
      {Lef. /Reg.} & $1.08$ & $1.12$ & $3.44$ & -- & -- \\ 
      \hline 
    \end{tabular}

  \caption{Timings comparison (in sec.) of different implementations of the polynomial approximation generation and of Lefèvre and regular HR-case searches in $[1, 2[$.}
  \label{table_perf_hr_case_1_2}
\end{table}

\begin{table*}
  \centering

  \begin{tabular}{ccccccc}
    \cline{2-7}
    \multicolumn{1}{c}{} & \multicolumn{2}{c}{MPI} & \multicolumn{2}{c}{CPU-GPU} & \multicolumn{2}{c}{MPI Lef./CPU-GPU}\\
    \multicolumn{1}{c}{}& { $[1,2[$} &  $[128,256[$ & { $[1,2[$} &  $[128,256[$ & { $[1,2[$}&  $[128,256[$ \\
    \hline
    Polynomial approximation generation & $5336.81$ & $11243.26$ & $785.14$ & $1612.03$ & $6.74$ & $6.97$ \\
    Lefèvre HR-case search & $5292.67$ & $169911.90$ & $2446.78$ & $51530.44$ & $2.16$ & $3.30$\\
    Regular HR-case search & $4716.96$ & -- & $711.8$ & $61581.87$  & $7.44$  & $2.76$\\
    \hline
  \end{tabular}

  \caption{Timings (in seconds) for binades $[1,2[$ and $[128,256[$. Timings for the MPI regular HR-case search over $[128,256[$ have been omitted because they are prohibitive.}
  \label{table_timings}
\end{table*}

We show in Table \ref{table_perf_hr_case_1_2} performance results of the HR-case
search over the binade $[1,2[$ as it corresponds to the general case according
to \cite{Lefevre2005}. First, we remark that Lefèvre and the regular HR-case
searches take advantage of the two-way SMT on the multi-core tests as we have a
parallel speedup higher than the number of cores.
Then, the deployment of Lefèvre HR-case search on GPU offers a good speedup of
15.05x over one CPU core and 2.16x over six cores.  Finally, the new regular
HR-case search delivers over Lefèvre HR-case search a slight gain of 8\% on
one CPU core, of 12\% over six CPU cores and an important speedup of 3.44x on
GPU. This result in a very good speedup of 51.71x for the regular HR-case search
on GPU over Lefèvre HR-case search on one CPU core, and of 7.43x over six CPU
cores.

\subsection{Polynomial approximation generation}

We show in Table \ref{table_perf_hr_case_1_2} performance results of the
polynomial approximation generation step over the binade $[1,2[$.  We first
observe that the polynomial approximation generation takes great advantage of
SMT with a speedup of 8.25x when using six CPU cores. This is mainly due to the
high latency caused by the carry propagation during the multi-precision addition
which can be partly offset by the SMT execution.  Concerning the CPU-GPU
deployment of the polynomial approximation generation, the times includes the
CPU computations, the data transfers from CPU to GPU and the GPU
computation. This hybrid CPU-GPU deployment greatly takes advantage of the GPU
as all the threads perform independent computations and as the control flow is
perfectly regular among the GPU threads. It offers thus a speedup of 54.89x over
the one CPU core execution and of 6.66x over the six core execution.

\subsection{Overall performance results}

In this subsection, we present detailed performance results for the overall
algorithms on different binades. In the following tables, one can remark that
the total times are slightly higher than the sum of the three phases. This is
due to the cost of measuring time for each phase. In Table \ref{table_timings},
the timings are obtained over two binades. The binade $[1,2[$ corresponds to the
general case according to \cite{Lefevre2005}, where the \emph{$\exp{}$} function
is well approximated by a polynomial of degree one, and the binade $[128,256[$
corresponds to the last entire binade before overflow, where the \emph{$\exp{}$}
function is hard to approximate by a polynomial of degree one.

We can first observe that speedups on CPU-GPU over CPU of the polynomial
approximation generation are similar, even if in the binade $[128,256[$ we use
longer multi-precision words (maximum coefficient sizes are 320 bits for the
binade $[1,2[$ and 448 for the binade $[128,256[$) and polynomials of higher
degree ($\max_j{(\operatorname{deg} r_{t,j}(x))}$ is 6 for the binade $[1,2[$
and 10 for the binade $[128,256[$).

It has to be noticed that the HR-case search is much slower in the binade
$[128,256[$ than in the binade $[1,2[$ (22.7x and 86.5x for Lefèvre and regular
HR-case searches respectively on GPU). Moreover, Lefèvre HR-case search delivers
a better speedup on GPU over CPU in $[128,256[$ compared to $[1,2[$, and regular
HR-case search delivers a lower speedup. The high computation times required in
the binade $[128,256[$ and the disparities in speedups of Lefèvre and regular
HR-case searches can be explained by the truncation error $\epsilon_{trunc}$
introduced by the Boolean tests used in both filtering strategies.


\begin{table}
  \centering
    \begin{tabular}{ccccc}
      \cline{2-5}
      \multicolumn{1}{r}{} & \multicolumn{2}{c}{Lefèvre} & \multicolumn{2}{c}{Regular}\\
      \multicolumn{1}{r}{} & Arguments & Time (s) & Arguments & Time (s)\\
      \hline
      {Phase 1}& $9.01\cdot 10^{15}$ & $2372.60$ & $9.01\cdot 10^{15}$ & $583.97$\\
      {Phase 2} & $3.19\cdot 10^{13}$ & $61.31$ & $1.62\cdot 10^{14}$ & $91.41$\\
      {Phase 3} & $7.65\cdot 10^{10}$ & $11.02$ & $5.14\cdot 10^{11}$ & $35.17$\\
      \hline
    \end{tabular}
    \caption{Details on each phase for Lefèvre and regular HR-case searches on GPU in the binade $[1,2[$.}
    \label{table_phase_timing_1_2}
\end{table}

We therefore present in Table \ref{table_phase_timing_1_2} the filtering and
timing details of the Lefèvre and the regular HR-case searches over the binade
$[1,2[$. In Table \ref{table_phase_timing_1_2}, both HR-case searches split the
entering intervals into 8 sub-intervals in phase 2.  In this binade the
\emph{$\exp{}$} function is well approximated by a polynomial of degree
one. This implies that the error due to the truncation to degree one is low
compared to the error $\epsilon$ we want to test, and the Boolean tests used in
Lefèvre and regular HR-case searches fail rarely.

However, as stated in Sect. \ref{sect_filtering}, the new regular HR-case search
filters less intervals than Lefèvre HR-case search. This increases the amount of
time spent in phases 2 and 3 by a factor 1.49 and 3.19 respectively.
Nevertheless, we can observe a good speedup of 4.06x in phase 1 due to the
regularity of the new regular HR-case search. As phases 2 and 3 are minority,
the new regular HR-case search offers a total speedup of 3.44x over Lefèvre
HR-case search.


\begin{table}
  \centering
    \begin{tabular}{ccccc}
      \cline{2-5}
      \multicolumn{1}{r}{} & \multicolumn{2}{c}{Lefèvre} & \multicolumn{2}{c}{Regular} \\
      \multicolumn{1}{r}{} & Arguments & Time (s) & Arguments & Time (s)\\
      \hline
      {Phase 1}& $9.01\cdot 10^{15}$ & $4097.41$ & $9.01\cdot 10^{15}$ & $1634.67$\\
      {Phase 2} & $8.97\cdot 10^{15}$ & $30003.78$ & $9.01\cdot 10^{15}$ & $21443.58$\\
      {Phase 3} & $4.19 \cdot 10^{14}$ & $17428.16$ & $9.00\cdot 10^{14}$ & $38480.87$\\
      \hline
    \end{tabular}
    \caption{Details on each phase for Lefèvre and regular HR-case searches on GPU in the binade $[128,256[$.}
    \label{table_phase_timing_128_256}
\end{table}


Table \ref{table_phase_timing_128_256} details the corresponding results for the
binade $[128,256[$ where the \emph{$\exp{}$} function is hard to approximate by
a polynomial of degree one. This implies that $\epsilon_{trunc}$ is high
compared to $\epsilon$, and the Boolean tests used in Lefèvre and regular
HR-case searches fail very often. We set the Lefèvre HR-case search to split the
entering intervals into 16 sub-intervals in phase 2 and the regular HR-case
search to split the entering intervals into 32 sub-intervals in phase 2. This is
due to the need of balancing phase 2 and phase 3 in order to obtain the best
performance. Here, the regular HR-case search has to use parallel prefix sums
for the compaction operation between each phase since the Boolean tests fail
very often (see Sect. \ref{sect_filtering}).

This very high failure rate of the Boolean tests also implies that the critical
phases for this binade are the phases 2 and 3. Hence, Lefèvre HR-case search is
more efficient as it filters more than the new regular HR-case search. In this
binade, 10.00\% of the initial arguments are involved in phase 3 with the
regular HR-case search against 4.65\% with Lefèvre HR-case search. This results
in Lefèvre HR-case search being 16.3\% faster than the regular HR-case search on
GPU for this binade. Moreover, as phase 3 corresponds to the exhaustive search,
which is embarrassingly parallel and which offers a completely regular control
flow, we still have a good speedup on GPU (up to 3.30x with Lefèvre HR-case
search over a hex-core CPU).

Hence, both HR-case searches can be used depending on the truncation error. The
latter directly depends on the coefficient of the term of degree two of the
approximation polynomial. A threshold on the truncation error to switch from one
HR-case search to the other can be precomputed. One can also use the ratio of
the number of intervals in phase 3 over the number of intervals in phase 1 of
the previous interval to select the appropriate HR-case search algorithm for the
current interval. As shown in Table \ref{table_timings}, this let us benefit
from a very good speedup of 7.44x on a GPU over a hex-core CPU when the function
is well approximated by a polynomial of degree one, and from a good speedup of
3.30x otherwise. For example, with the exponential function in double precision,
out of twenty-two binades which do not evaluate to overflow or underflow,
fifteen binades are more efficiently computed using the new regular algorithm.

However, both HR-case searches are slow when the truncation error is high
compared to the targeted error. The best here should be to consider a
Boolean test using polynomials of higher degree like in the SLZ algorithm.


\section{Conclusion and future work}
\label{sect_ccl}

In this paper, we have proposed a new algorithm based on continued fraction
expansion for HR-case search which improves Lefèvre HR-case search algorithm by
strongly reducing loop and branch divergence, which is a problem inherent to GPU
because of their partial SIMD architecture. We have also proposed an efficient
deployment on GPU of these two HR-case search algorithms and an hybrid CPU-GPU
deployment for the generation of polynomial approximations.

When searching for HR-cases of the \emph{$\exp{}$} function in double precision,
these deployments enable an overall speedup of up to 53.4x on one GPU over a
sequential execution on one CPU core, and a speedup of up to 7.1x on one GPU
over one hex-core CPU.

In the future, we plan to investigate whether the regular HR-case search can
benefit from other SIMD architectures like vector units (SSE, AVX, $\dots$) on
multi-core CPU and Intel Xeon Phi architectures. This will require
an OpenCL \cite{OpenCL} implementation and an effective automatic vectorization
by the OpenCL compiler.

\alert{We also plan to provide formal proofs of the deployed algorithms, and
  certificates along with the produced hardness-to-round. This is eased by the
  continued fraction expansion formalism, and would enable a validated
  generation of hardness-to-round, which is necessary to improve the confidence
  in the produced results. This is necessary before computing the
  hardness-to-round of all the functions recommended by the IEEE standard 754.}

Finally, we hope to tackle the quadruple precision by deploying on GPU the SLZ
algorithm which tests the existence of HR-cases with higher degree
polynomials. This algorithm heavily relies on the use of the LLL algorithm. The
deployment of this algorithm on GPU is therefore far from trivial if one wants
to obtain good performance. Porting the LLL algorithm to GPU will be the next
step of this work.


\section{Acknowledgement}
This work was supported by the TaMaDi project of the French ANR (grant ANR 2010
BLAN 0203 01). The authors thank Vincent Lefèvre for helpful discussions, and
Polytech Paris-UPMC for the CPU-GPU server. Finally, they would like thank the
reviewers for helping them to improve the readability and the quality of the
paper.

\bibliographystyle{ieeetr} 
\bibliography{MyCollection}

\listoftodos

\end{document}